%% file: diluted_tfim.tex
\documentclass[aps,prb,twocolumn,nofootinbib]{revtex4-2}
\usepackage[utf8]{inputenc}
\usepackage[T1]{fontenc}

\usepackage{amsmath}
\usepackage{amsfonts}
\usepackage{amssymb}
\usepackage{graphicx}
\usepackage{dsfont}

\usepackage[english]{babel}

\usepackage{physics}

\usepackage{hyphenat}
\usepackage{hyperref}
\usepackage{xcolor}
\usepackage{xspace}

\newcommand{\ie}{i.\,e.\,}
\newcommand{\eg}{e.\,g.\,}

\begin{document}

\title{Quantum Monte Carlo study of the bond- and site-diluted transverse-field Ising model}
\author{Calvin Krämer}
\affiliation{Department Physik, Staudtstra{\ss}e 7, Universit\"at Erlangen-N\"urnberg, D-91058 Erlangen, Germany}
\author{Max Hörmann}
\affiliation{Department Physik, Staudtstra{\ss}e 7, Universit\"at Erlangen-N\"urnberg, D-91058 Erlangen, Germany}
\author{Kai Phillip Schmidt}
\affiliation{Department Physik, Staudtstra{\ss}e 7, Universit\"at Erlangen-N\"urnberg, D-91058 Erlangen, Germany}

\begin{abstract}
We study the transverse-field Ising model on a square lattice with bond- and site-dilution at zero temperature by stochastic series expansion quantum Monte Carlo simulations. 
Tuning the transverse field $h$ and the dilution $p$, the quantum phase diagram of both models is explored.
Both quantum phase diagrams show long-range order for small $h$ and small $p$. 
The ordered phase of each is separated from the disordered (quantum) Griffiths phase by second-order phase transitions on two critical lines touching at a multi-critical point.
Using Binder ratios we locate quantum critical points with high accuracy.
The order-parameter critical exponent $\beta$ and the average correlation-length exponent $\nu_{\mathrm{av}}$ are determined along the critical lines and at the multi-critical points via finite-size scaling.
We find three internally consistent sets of critical exponents and compare them with potentially connected universality classes.
The quantum Griffiths phase in the vicinity of the phase transition lines is analyzed through the local susceptibility.
Our results indicate that activated scaling occurs not only at the percolation transition, but also at the phase transition line for $p$ smaller than the percolation threshold $p_c$.
\end{abstract}

\maketitle

\section{Introduction}
Disorder plays a crucial role in shaping the physical properties of many condensed matter systems, often leading to emergent behavior different from that observed in clean, translationally invariant systems \cite{Anderson1958, Griffiths1969, Fisher1992, Vojta2019}.
Experimentally, disordered quantum magnets are realized on several platforms, \eg, by inserting non-magnetic ions, which effectively creates diluted, random-bond or random-field spin models \cite{Schechter2008,Stone2011,Babkevich2016, Manaka2008, Hong2010}.
In particular, disordered quantum systems can exhibit rich and subtle phenomena such as localisation \cite{Anderson1958, Thouless1974, Cardy1978, Basko2006, Oganesyan2007}, glassy dynamics \cite{Wu1991, Brooke1999, Rieger1994}, and rare-region effects \cite{Griffiths1969, McCoy1969, Fisher1992, Fisher1995}, which continue to challenge our understanding of quantum phase transitions in disordered systems. 
In low-dimensional systems, these effects are particularly pronounced.
In general, also the Harris criterion \cite{Harris1974} is more likely to be violated in low-dimensional systems giving rise to changes in the universality class.
Among the most striking disorder-induced phenomena are Griffiths-McCoy singularities \cite{Griffiths1969, McCoy1969, Bauer2005, Schroeder2011, Wang2017}, which arise in quantum systems with quenched randomness and are characterized by strong sample-to-sample fluctuations and slow dynamical responses near criticality.
\\
The transverse-field Ising model (TFIM) is a paradigmatic model for exploring quantum criticality in the presence of disorder \cite{Pfeuty1979, Young1996, Fisher1992, Rieger1998, Kovacs2010, Monthus2012, Igloi2018, Kraemer2024}. When dilution is introduced - either by randomly removing spins or by removing bonds - the interplay between quantum fluctuations and randomness gives rise to a rich quantum phase diagram \cite{Harris1974v2, Stinchcombe1981, Santos1982, Fittipaldi1985}, featuring (quantum) Griffiths phases \cite{Ikegami1998, Thompson2019, Kovcs2022} and unconventional critical behavior in the form of activated scaling \cite{Fisher1995, Senthil1996, Ikegami1998}.
\\
Refs.~\cite{Harris1974v2, Stinchcombe1981, Santos1982, Fittipaldi1985} find that long-range order still persists in these models as long as the transverse field is weak enough and the lattice percolates, \ie, if there is a connected cluster of spins that spans from one side of the system to the opposite side.
Large parts of the quantum phase diagrams are dominated by Griffiths singularities \cite{Senthil1996, Ikegami1998, Thompson2019}. 
Furthermore, Ref.~\cite{Senthil1996} related the scaling of observables at the percolation transition with those of the random transverse-field Ising chain, deducing that along this line activated dynamic scaling has to be present.
This was later numerically confirmed \cite{Ikegami1998}.
\\
In this work we quantitatively explore the quantum phase diagram of the ferromagnetic bond- and site-diluted TFIM on the square lattice.
We employ the stochastic series expansion (SSE) quantum Monte Carlo (QMC) method \cite{Sandvik1991,Sandvik1992,Sandvik2003,Sandvik2010} to simulate these models at effectively zero temperature on finite systems.
Utilizing finite-size scaling, we determine critical points along the phase boundary and the positions of the multi-critical points.
We also give estimates for the order-parameter critical exponents $\beta$ and the average correlation-length exponents $\nu_{\mathrm{av}}$ throughout the quantum phase diagram taking care of corrections to finite-size scaling.
Furthermore, we investigate the Griffiths phase close to the critical lines in order to detect activated scaling.
\\
The paper is organized as follows:
In Sec.~\ref{sec:model} we introduce the Hamiltonians and summarize existing results on the phases and phase transitions of the models.
The SSE quantum Monte Carlo method together with the observables and finite-size scaling forms utilized are presented in Sec.~\ref{sec:methods}.
Finally, in Sec.~\ref{sec:results} we present our results before concluding in Sec.~\ref{sec:conclusion}.

\section{Model}
\label{sec:model}
The Hamiltonians of the bond-diluted (BD) and site-diluted (SD) TFIMs are given by
\begin{align}
\mathcal{H}_{\mathrm{BD}} &=  J \sum_{\langle i j \rangle} \epsilon_{ij} \, \sigma^z_i \sigma^z_j - h \sum_i \sigma^x_i \label{eq:HBD} \\
\mathcal{H}_{\mathrm{SD}} &=  J \sum_{\langle i j \rangle} \epsilon_{i} \epsilon_{j} \, \sigma^z_i \sigma^z_j - h \sum_i \epsilon_{i} \,\sigma^x_i \label{eq:HSD}
\end{align}
respectively, where $\sigma^{x/z}_i$ are Pauli matrices acting on spins 1/2 arranged on a square lattice. 
The dilution is controlled by the parameter $\epsilon_{ij}$ ($\epsilon_{i}$). It is 0 with probability $p$ and 1 with probability $1-p$, \ie, it determines whether there is a bond between spins $i$ and $j$ for BD (a spin at site $i$ for SD) or not.
We set $J := -1$, \ie, consider ferromagnetic interactions and use $h$ and $p$ as control parameters.
\\
The quantum phase diagrams of these models have been explored in Refs.~\cite{Harris1974v2, Stinchcombe1981, Santos1982, Senthil1996, Ikegami1998, Thompson2019}.
In the limit $p=0$ the well-known clean TFIM on a square lattice is recovered. 
For small transverse fields $h$, the ground state exhibits long-range order with broken $\mathbb{Z}_2$-symmetry.
For large $h$ the ground state is in the featureless polarized phase. 
These two phases are separated by a second-order phase transition of 3D-Ising universality \cite{Pelissetto2002, Sachdev2007, Kos2016}. 
For $h=0$, one finds the same long-range ordered phase for small $p$ and disconnected clusters of spins for large $p$. 
The classical phase transition at the percolation threshold $p = p_c$ ($p_c = 0.5$ for bond-dilution \cite{Bollobs2005} and $p_c \approx 0.4073$ for site-dilution \cite{Newman2000}) is described by ordinary percolation theory and the corresponding set of critical exponents \cite{Nijs1979, Pearson1980, Sokolov1986, Stauffer2018}. 
For both $h > 0$ and $p>0$ the situation is less clear. 
Coming from the limit of the pure TFIM, the critical point $h_c(p)$ decreases with increasing dilution $p<p_c$ until it reaches a multi-critical point $h_c(p=p_c)$ with a discontinuous jump in the phase boundary \cite{Harris1974v2}.
The universality class of the phase transition on this line is expected to be the one of the two-dimensional random-bond TFIM \cite{Senthil1996, Thompson2019}.
However, in two dimensions it is under debate whether the random-bond TFIM is governed by an infinite-disorder fixed point (IDFP) \cite{Kovacs2010} or by a different universality class \cite{Choi2023}.
This issue arising due to various limitations in numerical studies was discussed in Ref.~\cite{Kraemer2024}.
Cranking up the magnetic field $h$ at the classical percolation transition leads to a critical line at $p=p_c$, which still shows the critical exponents of ordinary percolation \cite{Senthil1996}.
On this critical line, quantum fluctuation arising from the finite magnetic field are dangerously irrelevant and lead to activated dynamical scaling \cite{Senthil1996}.
Close to the critical lines, there is a Griffiths phase where rare regions, \ie, strongly-correlated finite clusters, lead to small energy gaps and singular behaviour (Griffiths singularities) \cite{Senthil1996, Ikegami1998, Griffiths1969}.
In the models Eqs.~\eqref{eq:HBD} and \eqref{eq:HSD} the Griffiths phase is expected to extend up to $h=h_c(p=0)$, the critical point of the pure TFIM.
At this point, all clusters are supposed to be polarized \cite{Thompson2019, Senthil1996, Ikegami1998, Kovcs2022}.
For a similar model, the site-diluted random TFIM, which has additionally randomness in $J$ and $h$, critical lines and exponents were determined using the strong-disorder renormalisation group (SDRG) method \cite{Kovacs2022}.
This method is to the best of our knowledge not applicable to the kind of disorder considered here.

\section{Methods}
\label{sec:methods}

\subsection{Stochastic series expansion}

In this work we use the stochastic series expansion (SSE) quantum Monte Carlo method \cite{Sandvik1991, Sandvik1992, Sandvik2003, Sandvik2010, Adelhardt2024} to simulate finite spin systems of the bond- and site-diluted TFIM. 
We follow the scheme for arbitrary TFIMs introduced by A. Sandvik in Ref.~\cite{Sandvik2003}. 
The SSE method uses a high-temperature expansion of the partition function and extends the configuration space in imaginary time. 
The Hamiltonian of the TFIM can be decomposed into a sum of operators
\begin{align}
\mathcal{H} = - \sum_{i = 1}^N \sum_{j = 0}^{i} \mathcal{H}_{i, j} + c \ ,
\label{eq:decomp}
\end{align}
where $N$ is the total number of spins and $c$ is an irrelevant constant offset. 
The operators $\mathcal{H}_{i, j}$ are defined following Ref.~\cite{Sandvik2003} with respect to the chosen computational basis $\{ | \alpha \rangle \} = \{ | \sigma^z_1, ... \sigma^z_N \rangle \}$:
\begin{align}
	\mathcal{H}_{0,0} = \mathds{1} \hspace{1.5cm} &\mathcal{H}_{i, 0} = h \, \sigma^x_i \\ 
	\mathcal{H}_{i, i} = h \hspace{1.5cm} &\mathcal{H}_{i, j} = |J| - J \, \sigma^z_i \sigma^z_j 
\end{align}
with $i, j \geq 1$ and $j < i$.
$\mathcal{H}_{i,j}$ is only a part of $\mathcal{H}$ when spin $i$ and $j$ are next-neighbours and $\epsilon_{ij}$ (or $\epsilon_i \epsilon_j$) is 1. 
In all other cases, $\mathcal{H}_{i,j} = 0$.
$\mathcal{H}_{0,0}$ is introduced for algorithmic reasons and is not part of $\mathcal{H}$.
Expanding the partition function in $\beta \mathcal{H}$ and inserting the decomposition of $\mathcal{H}$, leads to
\begin{align}
	Z &= \mathrm{Tr}(e^{-\beta \mathcal{H}}) \\
	&= \sum_{ \{ | \alpha \rangle  \} } \sum_{n = 0}^\infty \frac{\beta^n}{n!} \, \langle \alpha |\, \Bigl( \sum_{i = 1}^N \sum_{j = 0}^i \mathcal{H}_{i, j} \Bigr)^n \, | \alpha \rangle \\
	&= \sum_{ \{ | \alpha \rangle  \} } \sum_{n = 0}^\infty \sum_{\{ S_n \}} \frac{\beta^n}{n!} \, \langle \alpha | \, \prod_{l = 1}^n \mathcal{H}_{i(l), j(l)} \, | \alpha \rangle \\
	&= \sum_{\omega \in \Omega} \pi(\beta, \omega)\ .
\end{align}
Here $\{S_n\}$ denotes the set of all possible sequences $S_n$ of operator-products $\prod_{l=1}^n \mathcal{H}_{i(l), j(l)}$ with length $n$. 
The SSE configuration space is given by $\Omega =  \{ |\alpha \rangle \} \times \{ S_n \}$, where the additional dimension $\{ S_n \}$ can be related to an imaginary time.
By introducing a fixed-length scheme the sum over $n$ is truncated at length $\mathcal{L}$ and all sequences with $n < \mathcal{L}$ are padded with trivial operators $\mathcal{H}_{0,0}$.
A suitable $\mathcal{L}$ can be chosen dynamically during the equilibration such that the truncation error is exponentially small \cite{Sandvik2010}.
Starting with an initial configuration $\omega \in \Omega$, both the state and the sequence are constantly updated.
A full Monte Carlo step consists of two types of updates: 
In the diagonal update, only operators diagonal in $|\alpha \rangle$ are inserted and removed, which changes $S_n$ but leaves $|\alpha \rangle$ unchanged.
This is done based on the Metropolis-Hastings probabilities according to the weights $\pi(\beta, \omega)$.
The off-diagonal cluster update exchanges diagonal with off-diagonal operators changing both $|\alpha \rangle$ and $S_n$ and is unlike local updates still efficient close to critical points.
Both updates are presented in full detail in Ref.~\cite{Sandvik2003}.

\subsection{Convergence to zero temperature}
Since the SSE method is a finite-temperature approach, we have to perform the simulation at sufficiently low temperature $T = 1/\beta$ to effectively study quantum phase transitions at zero temperature.
We employ the beta doubling method introduced in Ref.~\cite{Sandvik2002} and make sure that any thermal fluctuations are negligible.
In this scheme the sampling temperature is gradually decreased by a factor of 2 in each step.
Simultaneously the sequence is copied and glued together such that it has twice the length. 
This is a good starting point for the simulation in the next step since $\mathcal{L} \sim \beta$.
Measuring observables during this procedure enables us to check convergence in temperature \cite{Sandvik2002}.
This has proven crucial especially for disordered systems where activated scaling occurs. 
Here, the finite-size energy gap between ground and excited states at the critical point is affected by exponential scaling making temperature convergence more challenging \cite{Kraemer2024}.

\subsection{Observables}

Diagonal observables are measured by acting on the state $|\alpha \rangle$ and averaging over their eigenvalues during the simulation.
The ordered and disordered phases in the phase diagrams of the models considered in this work can be distinguished by the averaged $z$-magnetisation $m$ equilibrated to effectively $T=0$.
Therefore, we mainly focus on determining moments of the $z$-magnetisation
\begin{align}
m^{2n} = \left\langle \frac{1}{N^{2n}} \Bigl( \sum_{i = 1}^{N} \sigma_i^z \Bigr)^{2n} \right\rangle_{\mathrm{MC}} \ ,
\end{align}
where $\langle \dots \rangle_{\mathrm{MC}}$ denotes the Monte Carlo average, \ie, the average of the observable sampled according to the weights $\pi(\beta, \omega)$.
Additionally, we measure the local susceptibility to study the Griffiths phase close to the critical lines, which can be calculated by
\begin{align}
\chi_{i,\mathrm{local}} &= \int_0^\beta  \langle \sigma^z_i(\tau)\sigma^z_i(0) \rangle \, d\tau \\
&= \left\langle \frac{\beta}{n(n+1)} \Biggl( \, n + \Biggl( \, \sum_{p=0}^{n-1} \sigma^z_{i,p} \Biggr)^2 \ \Biggr) \right\rangle_{\mathrm{MC}} \ ,
\label{eq:localchi}
\end{align}
where $\langle \dots \rangle$ is the thermal average and $\sigma_{i, p}^z$ is the spin state measured at propagation step $p$ in the sequence iterating only over the $n$ non-trivial operators \cite{Sandvik2010}.
\\
We further have to average over a large amount of disorder realisations in order to extract averaged quantities.
The disorder average is denoted by $\left[ \dots \right]$. 
We keep the number of Monte Carlo steps in our simulations as small as possible in favor of more disorder realisations in the same time.
This is advantageous since the additional average over many disorder realisations compensates for the poor Monte Carlo averages of each single realisation \cite{Sandvik2002, Kraemer2024}.

\subsection{Finite-size scaling}
\begin{figure}
\centering
\includegraphics{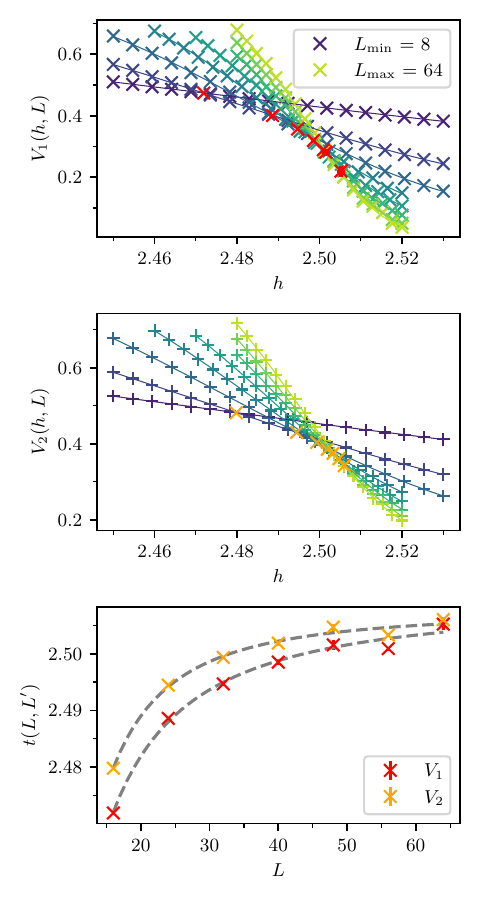}
\caption{Binder ratios $V_1$ (top) and $V_2$ (middle) and their intersections for the bond-diluted TFIM at $p=0.2$. To obtain an estimate for the critical point in the thermodynamic limit, the intersection points $t(L,L')$ of neighboring system sizes $L$ and $L'<L$ are extrapolated to $L=\infty$ (bottom).}
\label{fig:binder_example}
\end{figure}

In this work we want to extract critical points and critical exponents from observables measured on finite systems.
Finite-size scaling \cite{Wilson1970, Wilson1971, Hankey1972, Brzin1982, Brzin1985, Binder1987, Fisher1972, Kirkpatrick2015} connects the behavior of finite systems close to a continuous phase transition with the critical exponents of the infinite system.
The finite-size scaling form of the magnetisation $m$ is given by
\begin{align}
	m(r,L) = L^{-\beta/\nu} f_{m}(rL^{1/\nu}) \ ,
	\label{eq:fss_form}
\end{align}
where $r$ is the control parameter, $L$ the linear system size, $\beta$ the critical exponent of the order parameter, $\nu$ the correlation-length exponent and $f_{m}$ the scaling function.
In the context of the models considered in this work $r$ is either $(h-h_c)$ or $(p-p_c)$.
Furthermore, $\nu$ in the form above is replaced by $\nu_{\mathrm{av}}$, the average correlation-length exponent, since we scale disorder-averaged magnetisations $\left[ m(r,L)^{2n} \right]$.
Only the leading scaling of $m$ at the critical point is captured by Eq.~\eqref{eq:fss_form}.
In addition, there are corrections to scaling, whose specific form is usually not known and which are therefore hard to tackle \cite{Binder1987}.
In disordered systems, these corrections turned out to be quite significant \cite{Young1996, Kovacs2010, Lin2017, Choi2023, Kraemer2024}.
We therefore limit ourselves to determining $\beta/\nu_{\mathrm{av}}$ and $\nu_{\mathrm{av}}$ individually from $\left[ m(r=0,L)^{2n} \right]$, which has proven to be fruitful in order to tackle effects of corrections to scaling \cite{Kraemer2024}.
However, this requires prior knowledge of the position of the critical point ($h_c$ or $p_c$ respectively).
We determine the critical points via the intersections of Binder ratios \cite{Binder1987}
\begin{align}
V(r,L) = \frac{1}{2} \left(3 - \frac{  m^4(r,L) }{  m^2(r,L)^2} \right)
\end{align}
defined by moments of the $z$-magnetisation.
Since the leading scaling powers $L^{-\beta/\nu}$ of the moments of the magnetisation cancel
\begin{align}
V(r,L) = f_{V}(rL^{1/\nu}) 
\label{eq:scaling_binder}
\end{align}
has no scaling power and is therefore  independent of $L$ at the critical point $r=0$.
However, in the case of disorder, there is an additional degree of freedom in the definition of the Binder ratios, namely at which point the disorder average is taken.
In this work we consider the definitions
\begin{align}
V_1(r,L) &= \frac{1}{2} \left( 3 - \frac{\left[ m^4(r,L) \right]}{\left[ m^2(r,L) \right]^2} \right) \ , \\
V_2(r,L) &= \frac{1}{2} \left[ \left( 3 - \frac{ m^4(r,L)}{ m^2(r,L)^2} \right)  \right] \ ,
\end{align}
which have been reliable in the case of the random transverse-field Ising model \cite{Kraemer2024}.
Note that $V_1$ and $V_2$ would be exactly equal in the case of a pure system, but differ in the presence of disorder (see Fig.~\ref{fig:binder_example}).

\section{Results}
\label{sec:results}

In this section we first present the quantum phase diagram of the bond- and site-diluted TFIM obtained by the intersections of Binder ratios.
Then we focus on the critical exponents $\beta$ and $\nu_{\mathrm{av}}$ extracted from the scaling of the order parameter at the critical point.
Finally the Griffiths phase close to the phase boundaries is investigated using distributions of the local susceptibility.
We consider linear system sizes of up to $L=64$, \ie, 4096 spins.
We use the beta doubling method described above to check for each system size that thermal fluctuations are negligible.
Temperature convergence turned out to vary throughout the phase diagram. 
While a temperature of $\beta = 1/T = 2^{5}$ is sufficient for small $h$ at the percolation threshold $p=p_c$, we need temperatures of the order of $\beta = 2^{10}$ coming from the limit of the pure TFIM for $p<p_c$. 
Close to the multi-critical point ($h = h_c(p_c), p=p_c$), temperatures of up to $\beta = 2^{14}$ are necessary. 
The number of disorder realisations per data point ranges from a few hundred to tens of thousands depending on the inverse temperature $\beta$ and the system size $L$.
Raw data as well as processed data are provided in Ref.~\cite{RawData}.

\subsection{Quantum phase diagram}
\label{sec:phase_diagram}
\begin{figure}
\centering
\includegraphics{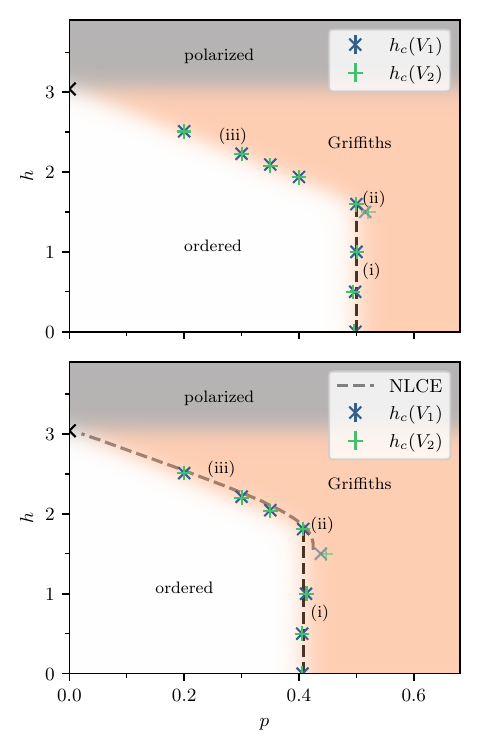}
\caption{Quantum phase diagrams of the bond-diluted (top) and site-diluted (bottom) TFIM. Critical points are obtained by the intersections of Binder ratios $V_1$ and $V_2$. The black dotted line is located at the percolation transition, \ie, at $p_c=0.5$ for bond-dilution \cite{Bollobs2005} and $p_c \approx 0.4073$ for site-dilution \cite{Newman2000}. The grey dotted line in the lower plot shows the phase boundary obtained by NLCE in Ref.~\cite{Thompson2019} for comparison. We group the phase transition lines into three regimes: (i) Vertical phase boundary at ($p=p_c, h<h_c(p_c)$), (ii) multi-critical point at ($h=h_c(p_c), p=p_c$), and (iii) phase boundary for $p<p_c$.}
\label{fig:phase_diagram}
\end{figure}

To determine the phase boundary of the ordered phase, we utilize the Binder ratios $V_1$ and $V_2$ (see Fig.~\ref{fig:binder_example}). 
Since Binder ratios are independent of $L$ at the critical point (see Eq.~\eqref{eq:scaling_binder}), the intersections of $V_{1/2}(r,L)$ are used to determine the critical point.
Due to corrections to scaling, which are not included in Eq.~\ref{eq:fss_form}, the $V_{1/2}(r,L)$ do not intersect at one point, but the intersections are shifted away from $r=0$.
To determine the position of the critical point in the thermodynamic limit we take the intersections $t(L,L')$ of the Binder ratios of systems of linear size $L$ with the one of the next smaller system size $L'$.
$t(L,L')$ is then extrapolated using algebraic fits (see Fig.~\ref{fig:binder_example} (bottom) for the $p=0.2$ bond-diluted model, the behaviour is similar for different $p$ as well as for site-dilution).
In Fig.~\ref{fig:binder_example} one can see, that although the same data is used for both $V_1$ and $V_2$, they do not intersect at the same positions.
The choice of where to take the disorder average $[\dots]$ is apparently not just a numerical subtlety, but leads to different sub-leading scaling corrections.
However, in the thermodynamic limit, both seem to agree on the position of the critical point.
\\ \\
The quantum phase diagrams of the bond- and site-diluted TFIM are shown in Fig.~\ref{fig:phase_diagram}.
Both exhibit three phases each:
The ordered phase, which is separated from the others by continuous phase transitions, the Griffiths phase, and the fully polarized phase.
The boundary between the latter two is defined by the field strength $h=h_c(p=0)$, where rare regions can no longer be ordered and therefore no longer lead to Griffiths singularities \cite{Thompson2019, Senthil1996, Ikegami1998}.
In this work, we did not investigate this boundary.
In the following, we divide the phase boundary between the ordered and disordered Griffiths phase into three regimes (see Fig.~\ref{fig:phase_diagram}): Regime (i) is the vertical critical line at $p=p_c$ for $h<h_c(p_c)$. Regime (ii) is the multi-critical point at ($h = h_c(p_c), p=p_c$). Regime (iii) denotes the critical line for $0<p<p_c$.
In Fig.~\ref{fig:phase_diagram} we show critical points in all three regimes determined both via intersections of $V_1$ and $V_2$.
Data points in regime (i), \ie, at the percolations threshold $p=p_c$ (black dashed lines in Fig.~\ref{fig:phase_diagram}) are determined by fixing $h$ and using $r=(p-p_c)$ as the control parameter.
We find a vertical phase boundary at $p=p_c$ as suggested by Refs.~\cite{Harris1974v2,Santos1982,Senthil1996,Ikegami1998}
Coming from the limit $h=0$ along the vertical phase boundary (i), corrections to finite-size scaling seemed to be weak, such that no extrapolation of intersections of Binder ratios was necessary.
However, approaching the multi-critical point (ii), quantum fluctuations become more and more important leading to stronger corrections.
Here, smaller temperatures were required in our simulation and strong, partly even non-monotonic scaling appeared at the intersections of Binder ratios.
Therefore, not knowing the functional dependence of the subleading finite-size corrections, the faded data points for $h=1.5$ in both phase diagrams were not extrapolated, but are just a mean-value of $t(L,L')$.
This leads clearly to an overestimation of the correct critical point $p_c$.
For data points at the critical line (iii) we use $r=(h-h_c)$ as control parameter at a fixed $p$.
As soon as $p>0$, corrections to scaling are present.
We extrapolate $t(L,L')$ here as shown in Fig.~\ref{fig:binder_example}.
Connecting our data points leads to an almost linear phase boundary (iii) between the critical point of the pure model at $p=0$ and the multi-critical point (ii).
We verified the consistency of our method by fixing $h = h_c(p)$ and determined the corresponding $p_c(h)$ using $r=(p-p_c)$ as control parameter.
The position of the multi-critical point (ii) is determined the same way by fixing $p=p_c$ and using $r=(h-h_c)$ as the control parameter.
We find $h_c(p_c)=1.597(4)$ for the bond-diluted TFIM and $h_c(p_c)=1.811(10)$ for the site-diluted TFIM.
The phase boundary of the site-diluted model agrees nicely with the results of Ref.~\cite{Thompson2019}, which investigated the same phase diagram using numerical linked cluster expansions (NLCE) (see grey dashed line in the lower panel of Fig.~\ref{fig:phase_diagram}).
A comparison with the phase boundaries determined in Ref.~\cite{Ikegami1998} is hardly possible since they focused on the finite-temperature phase diagram.

\subsection{Critical exponents}
\begin{figure*}
\centering
\includegraphics{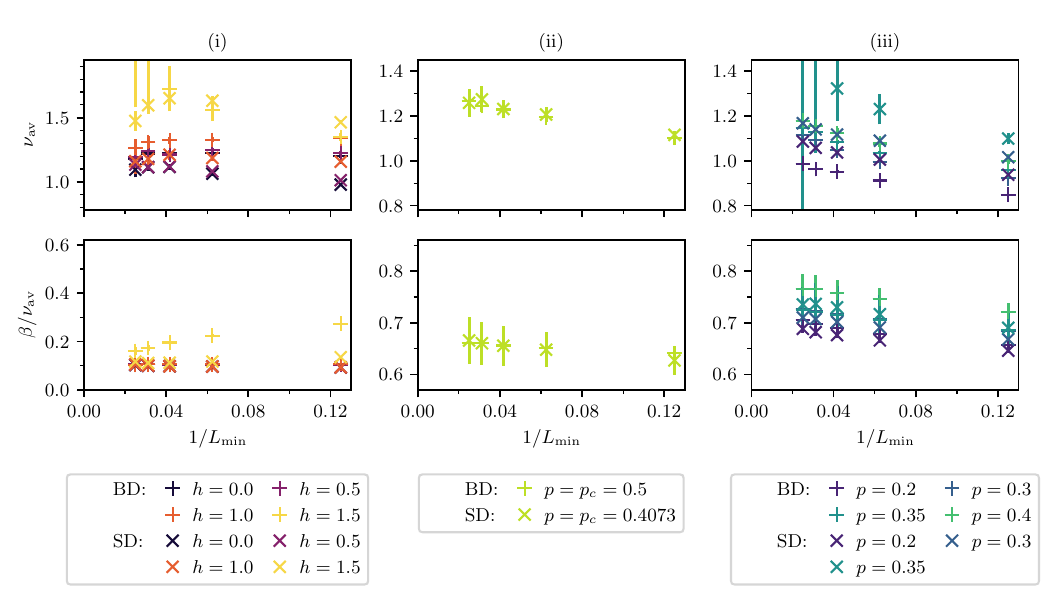}
\caption{Critical exponents $\nu_{\mathrm{av}}$ (top) and $\beta/\nu_{\mathrm{av}}$ (bottom) for the bond-diluted (BD) and site-diluted (SD) TFIM. The exponents are grouped in the three regimes (i), (ii) and (iii) (compare with Fig.~\ref{fig:phase_diagram}). (i) corresponds to the vertical phase boundary at ($p=p_c, h<h_c(p_c)$), (ii) is the multi-critical point at ($h=h_c(p_c), p=p_c$) and (iii) corresponds to critical points along the phase boundary for $p<p_c$. Smaller system sizes are successively removed from the fit to track the influence of corrections to finite-size scaling. $L_{\mathrm{min}}$ denotes the smallest system size included in the fit.}
\label{fig:all_exponents}
\end{figure*}
\begin{table*}
\centering
\caption{Critical exponents $\beta/\nu_{\mathrm{av}}$ and $\nu_{\mathrm{av}}$ compared with literature values for the 2D-IDFP determined by SDRG \cite{Kovacs2010} and other QMC studies on the 2D random-bond TFIM \cite{Choi2023,Kraemer2024}. The second to last and last column show the exponents from this work for the diluted TFIM in regime (iii) ($p<p_c$) and regime (ii) ($p=p_c$, \ie, at the multi-critical point). The critical exponents determined by successively removing smaller system sizes (see Fig.~\ref{fig:all_exponents}) were extrapolated to $1/L_\mathrm{min} \rightarrow 0$.}
\begin{tabular}{c|c|c|c|c|c}
 & IDFP \cite{Kovacs2010} & Random-bond TFIM \cite{Choi2023} & Random-bond TFIM \cite{Kraemer2024} & diluted TFIMs (regime (iii)) & diluted TFIMs (regime (ii)) \\ \hline
$\beta/\nu_{\mathrm{av}}$ &   0.982(15) & 0.62(20)  &  0.663(12)\footnote{To compare on equal footing the exponents from Ref.~\cite{Kraemer2024} (Fig. 15) were extrapolated to $1/L_\mathrm{min} \rightarrow 0$ in the same way using the provided data in Ref.~\cite{zenodo_kraemer2024}.} & 0.736(10) &  0.668(4) \\
$\nu_{\mathrm{av}}$ & 1.24(2)  & 0.95(2)  & 0.962(19)\footnotemark[1] & 1.10(4) & 1.295(5)
\end{tabular}
\label{tab:compare_exponents}
\end{table*}
To determine the order-parameter critical exponent $\beta$ and the average correlation-length exponent $\nu_{\mathrm{av}}$, we investigate the order parameter $m^2$ at the critical point $r=0$.
As an estimate for the critical point we use the values obtained from the intersections of Binder ratios presented in Sec.~\ref{sec:phase_diagram}.
The scaling form of the squared magnetisation then reduces to
\begin{align}
\left[ m^2 (r = 0, L) \right] \sim L^{-2\beta/\nu_{\mathrm{av}}}  \ .
\label{eq:obs_beta}
\end{align}
To extract $\nu_{\mathrm{av}}$ we expand the scaling form of the magnetisation (see Eq.~\eqref{eq:fss_form}) close to $r=0$ and consider the following quantity:
\begin{align}
& \hspace{0.46cm} 1 - \left( \left[ m^2 (r = \delta) \right] / \left[ m^2  (r = 0) \right]  \right) \label{eq:obs_nu} \\
&= 1 - \frac{f_{m^2}(0) + \frac{\partial f_{m^2}}{\partial x}|_{x = 0} \delta  L^{1/\nu_{\mathrm{av}}} + O(\delta^2)}{f_{m^2}(0)} \\ &\sim  L^{1/\nu_{\mathrm{av}}} \ .
\end{align}
We obtain the critical exponents by fitting straight lines to the quantities defined in Eqs.~\eqref{eq:obs_beta} and \eqref{eq:obs_nu} in a double logarithmic plot over the system size $L$.
Furthermore, to detect the influence of corrections to scaling, we successively remove smaller system sizes $L<L_{\mathrm{min}}$ from the fit.
The critical exponents $\beta/\nu_{\mathrm{av}}$ and $\nu_{\mathrm{av}}$ for both models are depicted in Fig.~\ref{fig:all_exponents} split into the different regimes (i), (ii), and (iii) introduced in Fig.~\ref{fig:phase_diagram}.
The error bars in Fig.~\ref{fig:all_exponents} take into account the uncertainty of the estimates of the critical points with respect to the extrapolation and different definition of $V_{1/2}$.
Generally, we find that there is no systematic difference between bond- and site-dilution in each regime.
In regime (i) the critical exponents we find are well in line with those of the 2D ordinary percolation universality class: $\beta/\nu_{\mathrm{av}} = 5/36 \approx 0.104$ and $\nu_{\mathrm{av}} = 4/3$ \cite{Nijs1979, Pearson1980, Sokolov1986}.
With increasing field $h$ the corrections to scaling appear to get stronger and even show non-monotonic behaviour with $L_\mathrm{min}$.
For regime (ii) and (iii) we extrapolate the exponents to $1/L_\mathrm{min} \rightarrow 0$.
The values are given in Tab.~\ref{tab:compare_exponents} together with literature values for the 2D-IDFP obtained by SDRG \cite{Kovacs2010} and for the 2D random-bond TFIM obtained by QMC \cite{Choi2023,Kraemer2024}.
Comparing the values in Tab.~\ref{tab:compare_exponents} the conjecture is reasonable that the diluted TFIMs in regime (iii) are governed by the same universality class as the random-bond TFIM, as mentioned in Ref.~\cite{Senthil1996}.
$\beta/\nu_{\mathrm{av}}$ of the IDFP, on the other hand, seems to be less compatible with our estimate.
One must be careful that all data from this work and Refs.~\cite{Choi2023,Kraemer2024} were calculated numerically on finite systems and show strong finite-size effects.
Therefore, they may have a certain bias that goes beyond the statistical error even if corrections are taken into account empirically.
SDRG in 2D is also no longer exact in the thermodynamic limit, but must be carried out numerically on finite systems. 
However, in Ref.~\cite{Kovacs2010} they achieve much larger system sizes than we or Refs.~\cite{Choi2023,Kraemer2024} do using QMC.
It is therefore difficult to make a definitive statement as to whether the methods agree on the critical exponents of the random-bond and diluted TFIM in the thermodynamic limit.
\\
At the multi-critical point (regime (ii)) we find slightly different exponents compared to regime (iii) (compare last two columns in Tab.~\ref{tab:compare_exponents}).
Noticing that $\beta/\nu_{\mathrm{av}}$ shifts to slightly larger values with increasing $p$ in regime (iii), we suspect the exponents to change at the multi-critical point instead of observing a crossover effect between regime (i) and (iii).

\subsection{Griffiths phase}
\begin{figure}
\centering
\includegraphics{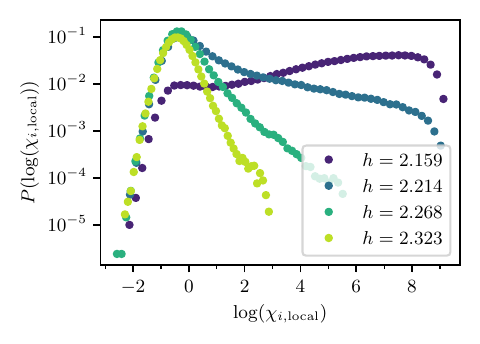}
\caption{Distribution of the local susceptibility $\chi_{i, \mathrm{local}}$ of the bond-diluted TFIM ($L=56$) at $p=0.3$ for different values of $h$. With increasing $h$ approaching the critical point ($h_c(p=0.3) \approx 2.23$) the slope of the tail is getting shallower, meaning $z'$ is increasing.}
\label{fig:localchi_example}
\end{figure}

In the Griffiths phase, rare regions have a strong impact on the low-energy physics.
Even though the formation of rare regions is exponentially suppressed, they still have an overall constant contribution to observables, since they exhibit exponentially small energy gaps \cite{Igloi2018}.
The distribution of energy gaps $\Delta$ for low energies scales as
\begin{align}
P(\Delta) \sim  \Delta^{\frac{d}{z'}-1} \ ,
\label{eq:dist_E}
\end{align}
where $d$ is the dimension of the system and $z'$ a pseudo-critical exponent \cite{Igloi2018}.
In analogy to the dynamical critical exponent $z$, $z'$ relates the length scales of the rare regions with their energy scales.
Nevertheless they cannot be identified with each other. 
In this work we do not directly determine the distribution of energy gaps, but instead use the local susceptibility $\chi_{i, \mathrm{local}}$ introduced in Eq.~\eqref{eq:localchi}.
The local susceptibility can be rewritten by evaluating the integral over imaginary time:
\begin{align}
\chi_{i, \mathrm{local}} &=  \int_0^\beta  \langle \sigma^z_i(\tau)\sigma^z_i(0) \rangle \, d\tau \nonumber \\
&=  \int_0^\beta \frac{1}{Z} \sum_{|\alpha \rangle, |\gamma \rangle} \langle \alpha | e^{H (\tau-\beta)} \sigma_i^z e^{-H \tau} |\gamma \rangle \langle \gamma| \sigma_i^z | \alpha \rangle  \, d\tau \nonumber \\
&= \frac{1}{Z}  \sum_{|\alpha \rangle \neq |\gamma \rangle} \frac{ |\langle \alpha |  \sigma_i^z | \gamma \rangle |^2}{E_\gamma-E_\alpha} \left( e^{-\beta E_\alpha} - e^{-\beta E_\gamma} \right) \nonumber \\
&= \frac{2}{Z} \sum_{|\alpha \rangle \neq |\gamma \rangle} \frac{|\langle \alpha |  \sigma_i^z | \gamma \rangle |^2}{E_\gamma-E_\alpha}  e^{-\beta E_\alpha} \ .
\end{align}
Here we assume that we still have a gapped spectrum.
Then, in the limit of small temperatures $T=1/\beta$, the dominant scaling of $\chi_{i, \mathrm{local}}$ is related to the energy gap between ground state and first excited state $\Delta$ as
\begin{align}
\chi_{i, \mathrm{local}}  \sim \frac{1}{\Delta}
\label{eq:dist_localchi}
\end{align}
neglecting the contributions of higher modes.
Whether this approximation is still valid directly at the critical point is questionable.
Similar to Eq.~\eqref{eq:dist_E} for small $\Delta$, we expect the distribution of the logarithm of $\chi_{i, \mathrm{local}}$ to scale as
\begin{align}
P(\log(\chi_{i, \mathrm{local}})) \sim  \left( \log(\chi_{i, \mathrm{local}}) \right)^{-\frac{d}{z'}}
\end{align}
for large $\chi_{i, \mathrm{local}}$.
The appearance of these exponential tails in the distribution of energy gaps or local susceptibilities respectively can be used to detect a Griffiths phase \cite{Young1996, Rieger1996, Rieger1998, Pich1998, Kraemer2024}.
We extract the exponent $d/z'$ by considering $P(\log \chi_{i, \mathrm{local}})$ on a log-scale, where the exponential tails appear as straight lines with slope $-d/z'$  (see Fig.~\ref{fig:localchi_example}):
\begin{align}
\log{\left(P(\log(\chi_{i, \mathrm{local}})) \right) } \sim -\frac{d}{z'} \log(\chi_{i, \mathrm{local}}) \  .
\end{align}
Furthermore, in systems obeying activated dynamic scaling, $z'$ diverges at the critical point since the relation between length and energy scales is no longer algebraically but exponentially at this point \cite{Rieger1996, Young1996, Pich1998, Fisher1995, Kraemer2024}.
\begin{figure}
\centering
\includegraphics{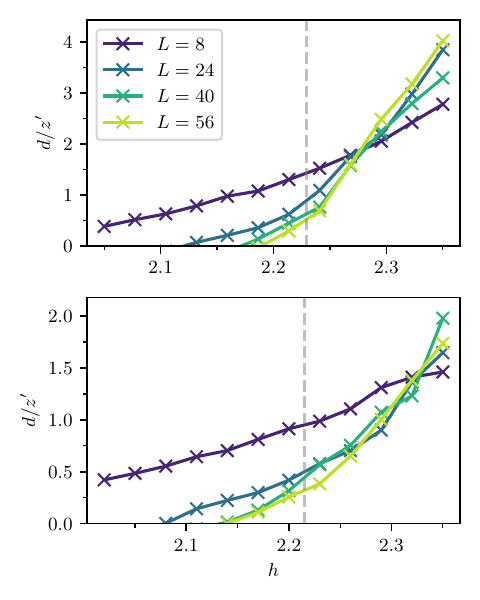}
\caption{$d/z'$ extracted at $p=0.3$ for the bond-diluted (top) and site-diluted (bottom) TFIM. For both models $z'$ diverges close to the critical point $h = h_c$ (indicated by dashed lines).}
\label{fig:dz_quantum}
\end{figure}
In Refs.~\cite{Senthil1996, Ikegami1998} it was shown that Griffiths singularities are present close to the percolation transition (regime (i) in Fig.~\ref{fig:phase_diagram}) and that activated scaling is present in this regime.
Our focus lies on the phase boundary for $p<p_c$ that was not fully captured in the previous works.
In Fig.~\ref{fig:dz_quantum} the exponents $d/z'$ are depicted for the bond- and site-diluted TFIM at $p=0.3$.
We can clearly see that for both models $z'$ diverges close to the critical point.
With increasing system size $L$ the divergence shifts towards the position of the critical point obtained from the Binder method above leading to a consistent picture within our methods.
This indicates that also along this phase boundary activated scaling is present.
For completeness we also looked at $\chi_{i, \mathrm{local}}$ for $h<h_c(p_c)$ close to the percolation transition and found similar exponential tails in $P(\log(\chi_{i, \mathrm{local}}))$ characterizing a Griffiths phase.
However, for small transverse fields $h$, we experienced issues extracting $d/z'$ close to the phase transition.
In Fig.~\ref{fig:dz_percolation} (bottom) one can see the exponent $d/z'$ for different $h<h_c(p_c)$ approaching the percolation transition.
More generally we find that $P(\log(\chi_{i, \mathrm{local}}))$ has a different form in the ordered phase for small $h$ than for large $h$ (compare $h=2.159$ curve in Fig.~\ref{fig:localchi_example} with $p=0.350$ curve in Fig.~\ref{fig:dz_percolation} (top)), which seems to also influence the form of the distribution close to the critical point.
In the ordered phase and especially at the critical point the assumption we made in Eq.~\eqref{eq:dist_localchi} are no longer justified.
Therefore different scaling that is not included in Eq.~\eqref{eq:dist_localchi} may contribute and dominate.
\begin{figure}
\centering
\includegraphics{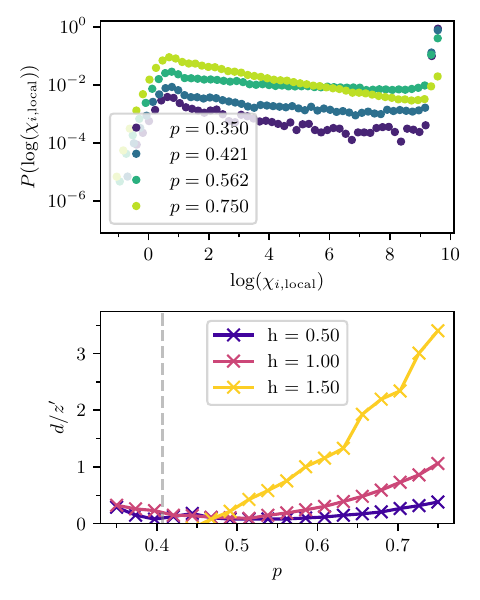}
\caption{Top: Distribution of the local susceptibility $\chi_{i, \mathrm{local}}$ of the site-diluted TFIM ($L=24$) at $h=0.5$ for different values of $p$. Bottom: $d/z'$ of the $L=24$ site-diluted TFIM extracted for different values of $h$ close to the percolation threshold $p_c$, indicated by the dashed line.}
\label{fig:dz_percolation}
\end{figure}
In Ref.~\cite{Ikegami1998} the cumulative distribution of $\chi_{i, \mathrm{local}}$ was used instead, presumably to overcome this issue.
This also stabilized our data for small $h$ (not shown here, we refer to Ref.~\cite{Ikegami1998}).

\section{Conclusion}
\label{sec:conclusion}
We investigate the quantum phase diagram of the bond- and site-diluted TFIM using a state-of-the-art quantum Monte Carlo method and a rigorous scheme to ensure that we only measure $T=0$ properties.
The phase boundaries for both models are determined with high accuracy using finite-size scaling, where we additionally take care of subleading corrections to finite-size scaling.
In particular we also determine the multi-critical points $h_c(p=p_c)$ for both types of dilution.
\\
We extract the order-parameter critical exponent $\beta$ as well as the average correlation-length exponent $\nu_{\mathrm{av}}$ for the first time from finite-size scaling.
Three regimes are identified:
Along the percolation threshold for zero and finite transverse field we find the critical exponents of ordinary percolation theory as expected from analytical considerations \cite{Senthil1996}.
At the phase boundary for $p<p_c$ the critical exponents are close to those obtained for the 2D random-bond TFIM in other quantum Monte Carlo studies \cite{Choi2023,Kraemer2024}.
The exponents at the multi-critical point seem to differ slightly from those at $p<p_c$.
However, we cannot fully rule out that they belong to the same universality class and whether they corresponds to an infinite disorder fixed point.
This remains an open question for future investigations.
\\
Finally, we find exponential tails in the distribution of the local susceptibility characteristic for Griffiths singularities.
The critical exponent $d/z'$ vanishes close to the phase boundaries indicating that activated scaling is present along the whole phase boundary.
Aiming towards more precise estimates for critical exponents, the origin and nature of corrections to finite-size scaling in disordered quantum systems have to be understood.
A rigorous theoretical framework is still needed to clarify how disorder alters scaling relations near quantum critical points.
However, our methods are still able to empirically capture corrections to scaling well and provide reliable estimates for quantum critical properties also in disordered systems.
As long as the model can be formulated free from the sign problem in the SSE quantum Monte Carlo framework, arbitrary variations to different spin models and extensions such as long-range interactions are feasible.

\section{Acknowledgements}
We thank J. A. Koziol and A. Langheld for valuable input and fruitful discussions.
We thankfully acknowledge the scientific support and HPC resources provided by the Erlangen National High Performance Computing Center (NHR@FAU) of the Friedrich-Alexander-Universität Erlangen-Nürnberg (FAU).

\subsection*{Funding information}
This work was funded by the Deutsche Forschungsgemeinschaft (DFG, German Research Foundation) - Project-ID 429529648 - TRR 306 QuCoLiMa (Quantum Cooperativity of Light and Matter). We acknowledge the support by the Munich Quantum Valley, which is supported by the Bavarian state government with funds from the Hightech Agenda Bayern Plus. The hardware of NHR@FAU is funded by the German Research Foundation DFG. 

\subsection*{Data availability}
The raw data used in this work as well as processed data are provided in Ref.~\cite{RawData}.

\input{diluted_tfim.bbl}

\end{document}

%% file: diluted_tfim.bbl
%

%% file: diluted_tfim.bbl
\begin{thebibliography}{67}%
\makeatletter
\providecommand \@ifxundefined [1]{%
 \@ifx{#1\undefined}
}%
\providecommand \@ifnum [1]{%
 \ifnum #1\expandafter \@firstoftwo
 \else \expandafter \@secondoftwo
 \fi
}%
\providecommand \@ifx [1]{%
 \ifx #1\expandafter \@firstoftwo
 \else \expandafter \@secondoftwo
 \fi
}%
\providecommand \natexlab [1]{#1}%
\providecommand \enquote  [1]{``#1''}%
\providecommand \bibnamefont  [1]{#1}%
\providecommand \bibfnamefont [1]{#1}%
\providecommand \citenamefont [1]{#1}%
\providecommand \href@noop [0]{\@secondoftwo}%
\providecommand \href [0]{\begingroup \@sanitize@url \@href}%
\providecommand \@href[1]{\@@startlink{#1}\@@href}%
\providecommand \@@href[1]{\endgroup#1\@@endlink}%
\providecommand \@sanitize@url [0]{\catcode `\\12\catcode `\$12\catcode
  `\&12\catcode `\#12\catcode `\^12\catcode `\_12\catcode `\%12\relax}%
\providecommand \@@startlink[1]{}%
\providecommand \@@endlink[0]{}%
\providecommand \url  [0]{\begingroup\@sanitize@url \@url }%
\providecommand \@url [1]{\endgroup\@href {#1}{\urlprefix }}%
\providecommand \urlprefix  [0]{URL }%
\providecommand \Eprint [0]{\href }%
\providecommand \doibase [0]{https://doi.org/}%
\providecommand \selectlanguage [0]{\@gobble}%
\providecommand \bibinfo  [0]{\@secondoftwo}%
\providecommand \bibfield  [0]{\@secondoftwo}%
\providecommand \translation [1]{[#1]}%
\providecommand \BibitemOpen [0]{}%
\providecommand \bibitemStop [0]{}%
\providecommand \bibitemNoStop [0]{.\EOS\space}%
\providecommand \EOS [0]{\spacefactor3000\relax}%
\providecommand \BibitemShut  [1]{\csname bibitem#1\endcsname}%
\let\auto@bib@innerbib\@empty
\bibitem [{\citenamefont {Anderson}(1958)}]{Anderson1958}%
  \BibitemOpen
  \bibfield  {author} {\bibinfo {author} {\bibfnamefont {P.~W.}\ \bibnamefont
  {Anderson}},\ }\bibfield  {title} {\bibinfo {title} {Absence of diffusion in
  certain random lattices},\ }\href {https://doi.org/10.1103/PhysRev.109.1492}
  {\bibfield  {journal} {\bibinfo  {journal} {Phys. Rev.}\ }\textbf {\bibinfo
  {volume} {109}},\ \bibinfo {pages} {1492} (\bibinfo {year}
  {1958})}\BibitemShut {NoStop}%
\bibitem [{\citenamefont {Griffiths}(1969)}]{Griffiths1969}%
  \BibitemOpen
  \bibfield  {author} {\bibinfo {author} {\bibfnamefont {R.~B.}\ \bibnamefont
  {Griffiths}},\ }\bibfield  {title} {\bibinfo {title} {Nonanalytic behavior
  above the critical point in a random ising ferromagnet},\ }\href
  {https://doi.org/10.1103/PhysRevLett.23.17} {\bibfield  {journal} {\bibinfo
  {journal} {Phys. Rev. Lett.}\ }\textbf {\bibinfo {volume} {23}},\ \bibinfo
  {pages} {17} (\bibinfo {year} {1969})}\BibitemShut {NoStop}%
\bibitem [{\citenamefont {Fisher}(1992)}]{Fisher1992}%
  \BibitemOpen
  \bibfield  {author} {\bibinfo {author} {\bibfnamefont {D.~S.}\ \bibnamefont
  {Fisher}},\ }\bibfield  {title} {\bibinfo {title} {Random transverse field
  ising spin chains},\ }\href {https://doi.org/10.1103/PhysRevLett.69.534}
  {\bibfield  {journal} {\bibinfo  {journal} {Phys. Rev. Lett.}\ }\textbf
  {\bibinfo {volume} {69}},\ \bibinfo {pages} {534} (\bibinfo {year}
  {1992})}\BibitemShut {NoStop}%
\bibitem [{\citenamefont {Vojta}(2019)}]{Vojta2019}%
  \BibitemOpen
  \bibfield  {author} {\bibinfo {author} {\bibfnamefont {T.}~\bibnamefont
  {Vojta}},\ }\bibfield  {title} {\bibinfo {title} {Disorder in quantum
  many-body systems},\ }\href
  {https://doi.org/10.1146/annurev-conmatphys-031218-013433} {\bibfield
  {journal} {\bibinfo  {journal} {Annual Review of Condensed Matter Physics}\
  }\textbf {\bibinfo {volume} {10}},\ \bibinfo {pages} {233} (\bibinfo {year}
  {2019})}\BibitemShut {NoStop}%
\bibitem [{\citenamefont {Schechter}(2008)}]{Schechter2008}%
  \BibitemOpen
  \bibfield  {author} {\bibinfo {author} {\bibfnamefont {M.}~\bibnamefont
  {Schechter}},\ }\bibfield  {title} {\bibinfo {title}
  {{${\mathrm{LiHo}}_{x}{\mathrm{Y}}_{1\ensuremath{-}x}{\mathrm{F}}_{4}$ as a
  random-field Ising ferromagnet}},\ }\bibfield  {journal} {\bibinfo  {journal}
  {Physical Review B}\ }\textbf {\bibinfo {volume} {77}},\ \href
  {https://doi.org/10.1103/physrevb.77.020401} {10.1103/physrevb.77.020401}
  (\bibinfo {year} {2008})\BibitemShut {NoStop}%
\bibitem [{\citenamefont {Stone}\ \emph {et~al.}(2011)\citenamefont {Stone},
  \citenamefont {Podlesnyak}, \citenamefont {Ehlers}, \citenamefont {Huq},
  \citenamefont {Samulon}, \citenamefont {Shapiro},\ and\ \citenamefont
  {Fisher}}]{Stone2011}%
  \BibitemOpen
  \bibfield  {author} {\bibinfo {author} {\bibfnamefont {M.~B.}\ \bibnamefont
  {Stone}}, \bibinfo {author} {\bibfnamefont {A.}~\bibnamefont {Podlesnyak}},
  \bibinfo {author} {\bibfnamefont {G.}~\bibnamefont {Ehlers}}, \bibinfo
  {author} {\bibfnamefont {A.}~\bibnamefont {Huq}}, \bibinfo {author}
  {\bibfnamefont {E.~C.}\ \bibnamefont {Samulon}}, \bibinfo {author}
  {\bibfnamefont {M.~C.}\ \bibnamefont {Shapiro}},\ and\ \bibinfo {author}
  {\bibfnamefont {I.~R.}\ \bibnamefont {Fisher}},\ }\bibfield  {title}
  {\bibinfo {title} {Persistence of magnons in a site-diluted dimerized
  frustrated antiferromagnet},\ }\href
  {https://doi.org/10.1088/0953-8984/23/41/416003} {\bibfield  {journal}
  {\bibinfo  {journal} {Journal of Physics: Condensed Matter}\ }\textbf
  {\bibinfo {volume} {23}},\ \bibinfo {pages} {416003} (\bibinfo {year}
  {2011})}\BibitemShut {NoStop}%
\bibitem [{\citenamefont {Babkevich}\ \emph {et~al.}(2016)\citenamefont
  {Babkevich}, \citenamefont {Nikseresht}, \citenamefont {Kovacevic},
  \citenamefont {Piatek}, \citenamefont {Dalla~Piazza}, \citenamefont
  {Kraemer}, \citenamefont {Kr\"{a}mer}, \citenamefont {Prokeš}, \citenamefont
  {Mat’aš}, \citenamefont {Jensen},\ and\ \citenamefont
  {Rønnow}}]{Babkevich2016}%
  \BibitemOpen
  \bibfield  {author} {\bibinfo {author} {\bibfnamefont {P.}~\bibnamefont
  {Babkevich}}, \bibinfo {author} {\bibfnamefont {N.}~\bibnamefont
  {Nikseresht}}, \bibinfo {author} {\bibfnamefont {I.}~\bibnamefont
  {Kovacevic}}, \bibinfo {author} {\bibfnamefont {J.~O.}\ \bibnamefont
  {Piatek}}, \bibinfo {author} {\bibfnamefont {B.}~\bibnamefont
  {Dalla~Piazza}}, \bibinfo {author} {\bibfnamefont {C.}~\bibnamefont
  {Kraemer}}, \bibinfo {author} {\bibfnamefont {K.~W.}\ \bibnamefont
  {Kr\"{a}mer}}, \bibinfo {author} {\bibfnamefont {K.}~\bibnamefont {Prokeš}},
  \bibinfo {author} {\bibfnamefont {S.}~\bibnamefont {Mat’aš}}, \bibinfo
  {author} {\bibfnamefont {J.}~\bibnamefont {Jensen}},\ and\ \bibinfo {author}
  {\bibfnamefont {H.~M.}\ \bibnamefont {Rønnow}},\ }\bibfield  {title}
  {\bibinfo {title} {{Phase diagram of diluted Ising ferromagnet
  ${\mathrm{LiHo}}_{x}{\mathrm{Y}}_{1\ensuremath{-}x}{\mathrm{F}}_{4}$}},\
  }\bibfield  {journal} {\bibinfo  {journal} {Physical Review B}\ }\textbf
  {\bibinfo {volume} {94}},\ \href {https://doi.org/10.1103/physrevb.94.174443}
  {10.1103/physrevb.94.174443} (\bibinfo {year} {2016})\BibitemShut {NoStop}%
\bibitem [{\citenamefont {Manaka}\ \emph {et~al.}(2008)\citenamefont {Manaka},
  \citenamefont {Kolomiets},\ and\ \citenamefont {Goto}}]{Manaka2008}%
  \BibitemOpen
  \bibfield  {author} {\bibinfo {author} {\bibfnamefont {H.}~\bibnamefont
  {Manaka}}, \bibinfo {author} {\bibfnamefont {A.~V.}\ \bibnamefont
  {Kolomiets}},\ and\ \bibinfo {author} {\bibfnamefont {T.}~\bibnamefont
  {Goto}},\ }\bibfield  {title} {\bibinfo {title} {{Disordered States in
  $\mathrm{IPA}\mathrm{\text{\ensuremath{-}}}\mathrm{Cu}({\mathrm{Cl}}_{x}{\mathrm{Br}}_{1\ensuremath{-}x}{)}_{3}$
  Induced by Bond Randomness}},\ }\bibfield  {journal} {\bibinfo  {journal}
  {Physical Review Letters}\ }\textbf {\bibinfo {volume} {101}},\ \href
  {https://doi.org/10.1103/physrevlett.101.077204}
  {10.1103/physrevlett.101.077204} (\bibinfo {year} {2008})\BibitemShut
  {NoStop}%
\bibitem [{\citenamefont {Hong}\ \emph {et~al.}(2010)\citenamefont {Hong},
  \citenamefont {Zheludev}, \citenamefont {Manaka},\ and\ \citenamefont
  {Regnault}}]{Hong2010}%
  \BibitemOpen
  \bibfield  {author} {\bibinfo {author} {\bibfnamefont {T.}~\bibnamefont
  {Hong}}, \bibinfo {author} {\bibfnamefont {A.}~\bibnamefont {Zheludev}},
  \bibinfo {author} {\bibfnamefont {H.}~\bibnamefont {Manaka}},\ and\ \bibinfo
  {author} {\bibfnamefont {L.-P.}\ \bibnamefont {Regnault}},\ }\bibfield
  {title} {\bibinfo {title} {{Evidence of a magnetic Bose glass in
  ${({\text{CH}}_{3})}_{2}{\text{CHNH}}_{3}\text{Cu}{({\text{Cl}}_{{0.95}}{\text{Br}}_{{0.05}})}_{3}$
  from neutron diffraction}},\ }\bibfield  {journal} {\bibinfo  {journal}
  {Physical Review B}\ }\textbf {\bibinfo {volume} {81}},\ \href
  {https://doi.org/10.1103/physrevb.81.060410} {10.1103/physrevb.81.060410}
  (\bibinfo {year} {2010})\BibitemShut {NoStop}%
\bibitem [{\citenamefont {Thouless}(1974)}]{Thouless1974}%
  \BibitemOpen
  \bibfield  {author} {\bibinfo {author} {\bibfnamefont {D.}~\bibnamefont
  {Thouless}},\ }\bibfield  {title} {\bibinfo {title} {Electrons in disordered
  systems and the theory of localization},\ }\href
  {https://doi.org/10.1016/0370-1573(74)90029-5} {\bibfield  {journal}
  {\bibinfo  {journal} {Physics Reports}\ }\textbf {\bibinfo {volume} {13}},\
  \bibinfo {pages} {93–142} (\bibinfo {year} {1974})}\BibitemShut {NoStop}%
\bibitem [{\citenamefont {Cardy}(1978)}]{Cardy1978}%
  \BibitemOpen
  \bibfield  {author} {\bibinfo {author} {\bibfnamefont {J.~L.}\ \bibnamefont
  {Cardy}},\ }\bibfield  {title} {\bibinfo {title} {Electron localisation in
  disordered systems and classical solutions in ginzburg-landau field theory},\
  }\href {https://doi.org/10.1088/0022-3719/11/8/006} {\bibfield  {journal}
  {\bibinfo  {journal} {Journal of Physics C: Solid State Physics}\ }\textbf
  {\bibinfo {volume} {11}},\ \bibinfo {pages} {L321–L327} (\bibinfo {year}
  {1978})}\BibitemShut {NoStop}%
\bibitem [{\citenamefont {Basko}\ \emph {et~al.}(2006)\citenamefont {Basko},
  \citenamefont {Aleiner},\ and\ \citenamefont {Altshuler}}]{Basko2006}%
  \BibitemOpen
  \bibfield  {author} {\bibinfo {author} {\bibfnamefont {D.}~\bibnamefont
  {Basko}}, \bibinfo {author} {\bibfnamefont {I.}~\bibnamefont {Aleiner}},\
  and\ \bibinfo {author} {\bibfnamefont {B.}~\bibnamefont {Altshuler}},\
  }\bibfield  {title} {\bibinfo {title} {Metal–insulator transition in a
  weakly interacting many-electron system with localized single-particle
  states},\ }\href {https://doi.org/10.1016/j.aop.2005.11.014} {\bibfield
  {journal} {\bibinfo  {journal} {Annals of Physics}\ }\textbf {\bibinfo
  {volume} {321}},\ \bibinfo {pages} {1126–1205} (\bibinfo {year}
  {2006})}\BibitemShut {NoStop}%
\bibitem [{\citenamefont {Oganesyan}\ and\ \citenamefont
  {Huse}(2007)}]{Oganesyan2007}%
  \BibitemOpen
  \bibfield  {author} {\bibinfo {author} {\bibfnamefont {V.}~\bibnamefont
  {Oganesyan}}\ and\ \bibinfo {author} {\bibfnamefont {D.~A.}\ \bibnamefont
  {Huse}},\ }\bibfield  {title} {\bibinfo {title} {Localization of interacting
  fermions at high temperature},\ }\bibfield  {journal} {\bibinfo  {journal}
  {Physical Review B}\ }\textbf {\bibinfo {volume} {75}},\ \href
  {https://doi.org/10.1103/physrevb.75.155111} {10.1103/physrevb.75.155111}
  (\bibinfo {year} {2007})\BibitemShut {NoStop}%
\bibitem [{\citenamefont {Wu}\ \emph {et~al.}(1991)\citenamefont {Wu},
  \citenamefont {Ellman}, \citenamefont {Rosenbaum}, \citenamefont {Aeppli},\
  and\ \citenamefont {Reich}}]{Wu1991}%
  \BibitemOpen
  \bibfield  {author} {\bibinfo {author} {\bibfnamefont {W.}~\bibnamefont
  {Wu}}, \bibinfo {author} {\bibfnamefont {B.}~\bibnamefont {Ellman}}, \bibinfo
  {author} {\bibfnamefont {T.}~\bibnamefont {Rosenbaum}}, \bibinfo {author}
  {\bibfnamefont {G.}~\bibnamefont {Aeppli}},\ and\ \bibinfo {author}
  {\bibfnamefont {D.}~\bibnamefont {Reich}},\ }\bibfield  {title} {\bibinfo
  {title} {From classical to quantum glass},\ }\href
  {https://doi.org/10.1103/physrevlett.67.2076} {\bibfield  {journal} {\bibinfo
   {journal} {Physical Review Letters}\ }\textbf {\bibinfo {volume} {67}},\
  \bibinfo {pages} {2076–2079} (\bibinfo {year} {1991})}\BibitemShut
  {NoStop}%
\bibitem [{\citenamefont {Brooke}\ \emph {et~al.}(1999)\citenamefont {Brooke},
  \citenamefont {Bitko}, \citenamefont {F.}, \citenamefont {Rosenbaum},\ and\
  \citenamefont {Aeppli}}]{Brooke1999}%
  \BibitemOpen
  \bibfield  {author} {\bibinfo {author} {\bibfnamefont {J.}~\bibnamefont
  {Brooke}}, \bibinfo {author} {\bibfnamefont {D.}~\bibnamefont {Bitko}},
  \bibinfo {author} {\bibfnamefont {T.}~\bibnamefont {F.}}, \bibinfo {author}
  {\bibnamefont {Rosenbaum}},\ and\ \bibinfo {author} {\bibfnamefont
  {G.}~\bibnamefont {Aeppli}},\ }\bibfield  {title} {\bibinfo {title} {Quantum
  annealing of a disordered magnet},\ }\href
  {https://doi.org/10.1126/science.284.5415.779} {\bibfield  {journal}
  {\bibinfo  {journal} {Science}\ }\textbf {\bibinfo {volume} {284}},\ \bibinfo
  {pages} {779–781} (\bibinfo {year} {1999})}\BibitemShut {NoStop}%
\bibitem [{\citenamefont {Rieger}\ and\ \citenamefont
  {Young}(1994)}]{Rieger1994}%
  \BibitemOpen
  \bibfield  {author} {\bibinfo {author} {\bibfnamefont {H.}~\bibnamefont
  {Rieger}}\ and\ \bibinfo {author} {\bibfnamefont {A.~P.}\ \bibnamefont
  {Young}},\ }\bibfield  {title} {\bibinfo {title} {Zero-temperature quantum
  phase transition of a two-dimensional ising spin glass},\ }\href
  {https://doi.org/10.1103/PhysRevLett.72.4141} {\bibfield  {journal} {\bibinfo
   {journal} {Phys. Rev. Lett.}\ }\textbf {\bibinfo {volume} {72}},\ \bibinfo
  {pages} {4141} (\bibinfo {year} {1994})}\BibitemShut {NoStop}%
\bibitem [{\citenamefont {McCoy}(1969)}]{McCoy1969}%
  \BibitemOpen
  \bibfield  {author} {\bibinfo {author} {\bibfnamefont {B.~M.}\ \bibnamefont
  {McCoy}},\ }\bibfield  {title} {\bibinfo {title} {Incompleteness of the
  critical exponent description for ferromagnetic systems containing random
  impurities},\ }\href {https://doi.org/10.1103/PhysRevLett.23.383} {\bibfield
  {journal} {\bibinfo  {journal} {Phys. Rev. Lett.}\ }\textbf {\bibinfo
  {volume} {23}},\ \bibinfo {pages} {383} (\bibinfo {year} {1969})}\BibitemShut
  {NoStop}%
\bibitem [{\citenamefont {Fisher}(1995)}]{Fisher1995}%
  \BibitemOpen
  \bibfield  {author} {\bibinfo {author} {\bibfnamefont {D.~S.}\ \bibnamefont
  {Fisher}},\ }\bibfield  {title} {\bibinfo {title} {Critical behavior of
  random transverse-field ising spin chains},\ }\href
  {https://doi.org/10.1103/PhysRevB.51.6411} {\bibfield  {journal} {\bibinfo
  {journal} {Phys. Rev. B}\ }\textbf {\bibinfo {volume} {51}},\ \bibinfo
  {pages} {6411} (\bibinfo {year} {1995})}\BibitemShut {NoStop}%
\bibitem [{\citenamefont {Harris}(1974{\natexlab{a}})}]{Harris1974}%
  \BibitemOpen
  \bibfield  {author} {\bibinfo {author} {\bibfnamefont {A.~B.}\ \bibnamefont
  {Harris}},\ }\bibfield  {title} {\bibinfo {title} {Effect of random defects
  on the critical behaviour of ising models},\ }\href
  {https://doi.org/10.1088/0022-3719/7/9/009} {\bibfield  {journal} {\bibinfo
  {journal} {Journal of Physics C: Solid State Physics}\ }\textbf {\bibinfo
  {volume} {7}},\ \bibinfo {pages} {1671} (\bibinfo {year}
  {1974}{\natexlab{a}})}\BibitemShut {NoStop}%
\bibitem [{\citenamefont {Bauer}\ \emph {et~al.}(2005)\citenamefont {Bauer},
  \citenamefont {Zapf}, \citenamefont {Ho}, \citenamefont {Butch},
  \citenamefont {Freeman}, \citenamefont {Sirvent},\ and\ \citenamefont
  {Maple}}]{Bauer2005}%
  \BibitemOpen
  \bibfield  {author} {\bibinfo {author} {\bibfnamefont {E.~D.}\ \bibnamefont
  {Bauer}}, \bibinfo {author} {\bibfnamefont {V.~S.}\ \bibnamefont {Zapf}},
  \bibinfo {author} {\bibfnamefont {P.-C.}\ \bibnamefont {Ho}}, \bibinfo
  {author} {\bibfnamefont {N.~P.}\ \bibnamefont {Butch}}, \bibinfo {author}
  {\bibfnamefont {E.~J.}\ \bibnamefont {Freeman}}, \bibinfo {author}
  {\bibfnamefont {C.}~\bibnamefont {Sirvent}},\ and\ \bibinfo {author}
  {\bibfnamefont {M.~B.}\ \bibnamefont {Maple}},\ }\bibfield  {title} {\bibinfo
  {title} {{Non-Fermi-Liquid Behavior within the Ferromagnetic Phase in
  ${\mathrm{U}\mathrm{R}\mathrm{u}}_{2\ensuremath{-}x}{\mathrm{R}\mathrm{e}}_{x}{\mathrm{S}\mathrm{i}}_{2}$}},\
  }\bibfield  {journal} {\bibinfo  {journal} {Physical Review Letters}\
  }\textbf {\bibinfo {volume} {94}},\ \href
  {https://doi.org/10.1103/physrevlett.94.046401}
  {10.1103/physrevlett.94.046401} (\bibinfo {year} {2005})\BibitemShut
  {NoStop}%
\bibitem [{\citenamefont {Schroeder}\ \emph {et~al.}(2011)\citenamefont
  {Schroeder}, \citenamefont {Ubaid-Kassis},\ and\ \citenamefont
  {Vojta}}]{Schroeder2011}%
  \BibitemOpen
  \bibfield  {author} {\bibinfo {author} {\bibfnamefont {A.}~\bibnamefont
  {Schroeder}}, \bibinfo {author} {\bibfnamefont {S.}~\bibnamefont
  {Ubaid-Kassis}},\ and\ \bibinfo {author} {\bibfnamefont {T.}~\bibnamefont
  {Vojta}},\ }\bibfield  {title} {\bibinfo {title} {Signatures of a quantum
  griffiths phase in a d-metal alloy close to its ferromagnetic quantum
  critical point},\ }\href {https://doi.org/10.1088/0953-8984/23/9/094205}
  {\bibfield  {journal} {\bibinfo  {journal} {Journal of Physics: Condensed
  Matter}\ }\textbf {\bibinfo {volume} {23}},\ \bibinfo {pages} {094205}
  (\bibinfo {year} {2011})}\BibitemShut {NoStop}%
\bibitem [{Wan(2017)}]{Wang2017}%
  \BibitemOpen
  \bibfield  {title} {\bibinfo {title} {{Quantum Griffiths Phase Inside the
  Ferromagnetic Phase of ${\mathrm{Ni}}_{1\ensuremath{-}x}{\mathrm{V}}_{x}$},
  author = {Wang, Ruizhe and Gebretsadik, Adane and Ubaid-Kassis, Sara and
  Schroeder, Almut and Vojta, Thomas and Baker, Peter J. and Pratt, Francis L.
  and Blundell, Stephen J. and Lancaster, Tom and Franke, Isabel and M\"oller,
  Johannes S. and Page, Katharine}},\ }\href
  {https://doi.org/10.1103/PhysRevLett.118.267202} {\bibfield  {journal}
  {\bibinfo  {journal} {Phys. Rev. Lett.}\ }\textbf {\bibinfo {volume} {118}},\
  \bibinfo {pages} {267202} (\bibinfo {year} {2017})}\BibitemShut {NoStop}%
\bibitem [{\citenamefont {Pfeuty}(1979)}]{Pfeuty1979}%
  \BibitemOpen
  \bibfield  {author} {\bibinfo {author} {\bibfnamefont {P.}~\bibnamefont
  {Pfeuty}},\ }\bibfield  {title} {\bibinfo {title} {An exact result for the 1d
  random ising model in a transverse field},\ }\href
  {https://doi.org/https://doi.org/10.1016/0375-9601(79)90017-3} {\bibfield
  {journal} {\bibinfo  {journal} {Physics Letters A}\ }\textbf {\bibinfo
  {volume} {72}},\ \bibinfo {pages} {245} (\bibinfo {year} {1979})}\BibitemShut
  {NoStop}%
\bibitem [{\citenamefont {Young}\ and\ \citenamefont
  {Rieger}(1996)}]{Young1996}%
  \BibitemOpen
  \bibfield  {author} {\bibinfo {author} {\bibfnamefont {A.~P.}\ \bibnamefont
  {Young}}\ and\ \bibinfo {author} {\bibfnamefont {H.}~\bibnamefont {Rieger}},\
  }\bibfield  {title} {\bibinfo {title} {Numerical study of the random
  transverse-field ising spin chain},\ }\href
  {https://doi.org/10.1103/PhysRevB.53.8486} {\bibfield  {journal} {\bibinfo
  {journal} {Phys. Rev. B}\ }\textbf {\bibinfo {volume} {53}},\ \bibinfo
  {pages} {8486} (\bibinfo {year} {1996})}\BibitemShut {NoStop}%
\bibitem [{\citenamefont {Rieger}\ and\ \citenamefont
  {Kawashima}(1999)}]{Rieger1998}%
  \BibitemOpen
  \bibfield  {author} {\bibinfo {author} {\bibfnamefont {H.}~\bibnamefont
  {Rieger}}\ and\ \bibinfo {author} {\bibfnamefont {N.}~\bibnamefont
  {Kawashima}},\ }\bibfield  {title} {\bibinfo {title} {Application of a
  continuous time cluster algorithm to the two-dimensional random quantum ising
  ferromagnet},\ }\href {https://doi.org/10.1007/s100510050761} {\bibfield
  {journal} {\bibinfo  {journal} {The European Physical Journal B - Condensed
  Matter and Complex Systems}\ }\textbf {\bibinfo {volume} {9}},\ \bibinfo
  {pages} {233} (\bibinfo {year} {1999})}\BibitemShut {NoStop}%
\bibitem [{\citenamefont {Kov\'acs}\ and\ \citenamefont
  {Igl\'oi}(2010)}]{Kovacs2010}%
  \BibitemOpen
  \bibfield  {author} {\bibinfo {author} {\bibfnamefont {I.~A.}\ \bibnamefont
  {Kov\'acs}}\ and\ \bibinfo {author} {\bibfnamefont {F.}~\bibnamefont
  {Igl\'oi}},\ }\bibfield  {title} {\bibinfo {title} {Renormalization group
  study of the two-dimensional random transverse-field ising model},\ }\href
  {https://doi.org/10.1103/PhysRevB.82.054437} {\bibfield  {journal} {\bibinfo
  {journal} {Phys. Rev. B}\ }\textbf {\bibinfo {volume} {82}},\ \bibinfo
  {pages} {054437} (\bibinfo {year} {2010})}\BibitemShut {NoStop}%
\bibitem [{\citenamefont {Monthus}\ and\ \citenamefont
  {Garel}(2012)}]{Monthus2012}%
  \BibitemOpen
  \bibfield  {author} {\bibinfo {author} {\bibfnamefont {C.}~\bibnamefont
  {Monthus}}\ and\ \bibinfo {author} {\bibfnamefont {T.}~\bibnamefont
  {Garel}},\ }\bibfield  {title} {\bibinfo {title} {The random transverse field
  ising model in d = 2: analysis via boundary strong disorder
  renormalization},\ }\href {https://doi.org/10.1088/1742-5468/2012/09/P09016}
  {\bibfield  {journal} {\bibinfo  {journal} {Journal of Statistical Mechanics:
  Theory and Experiment}\ }\textbf {\bibinfo {volume} {2012}},\ \bibinfo
  {pages} {P09016} (\bibinfo {year} {2012})}\BibitemShut {NoStop}%
\bibitem [{\citenamefont {Igl{\'{o}}i}\ and\ \citenamefont
  {Monthus}(2018)}]{Igloi2018}%
  \BibitemOpen
  \bibfield  {author} {\bibinfo {author} {\bibfnamefont {F.}~\bibnamefont
  {Igl{\'{o}}i}}\ and\ \bibinfo {author} {\bibfnamefont {C.}~\bibnamefont
  {Monthus}},\ }\bibfield  {title} {\bibinfo {title} {Strong disorder {RG}
  approach {\textendash} a short review of recent developments},\ }\bibfield
  {journal} {\bibinfo  {journal} {The European Physical Journal B}\ }\textbf
  {\bibinfo {volume} {91}},\ \href {https://doi.org/10.1140/epjb/e2018-90434-8}
  {10.1140/epjb/e2018-90434-8} (\bibinfo {year} {2018})\BibitemShut {NoStop}%
\bibitem [{\citenamefont {Krämer}\ \emph
  {et~al.}(2024{\natexlab{a}})\citenamefont {Krämer}, \citenamefont {Koziol},
  \citenamefont {Langheld}, \citenamefont {Hörmann},\ and\ \citenamefont
  {Schmidt}}]{Kraemer2024}%
  \BibitemOpen
  \bibfield  {author} {\bibinfo {author} {\bibfnamefont {C.}~\bibnamefont
  {Krämer}}, \bibinfo {author} {\bibfnamefont {J.~A.}\ \bibnamefont {Koziol}},
  \bibinfo {author} {\bibfnamefont {A.}~\bibnamefont {Langheld}}, \bibinfo
  {author} {\bibfnamefont {M.}~\bibnamefont {Hörmann}},\ and\ \bibinfo
  {author} {\bibfnamefont {K.~P.}\ \bibnamefont {Schmidt}},\ }\bibfield
  {title} {\bibinfo {title} {{Quantum-critical properties of the one- and
  two-dimensional random transverse-field Ising model from large-scale quantum
  Monte Carlo simulations}},\ }\href
  {https://doi.org/10.21468/SciPostPhys.17.2.061} {\bibfield  {journal}
  {\bibinfo  {journal} {SciPost Phys.}\ }\textbf {\bibinfo {volume} {17}},\
  \bibinfo {pages} {061} (\bibinfo {year} {2024}{\natexlab{a}})}\BibitemShut
  {NoStop}%
\bibitem [{\citenamefont {Harris}(1974{\natexlab{b}})}]{Harris1974v2}%
  \BibitemOpen
  \bibfield  {author} {\bibinfo {author} {\bibfnamefont {A.~B.}\ \bibnamefont
  {Harris}},\ }\bibfield  {title} {\bibinfo {title} {Upper bounds for the
  transition temperatures of generalized ising models},\ }\href
  {https://doi.org/10.1088/0022-3719/7/17/018} {\bibfield  {journal} {\bibinfo
  {journal} {Journal of Physics C: Solid State Physics}\ }\textbf {\bibinfo
  {volume} {7}},\ \bibinfo {pages} {3082–3102} (\bibinfo {year}
  {1974}{\natexlab{b}})}\BibitemShut {NoStop}%
\bibitem [{\citenamefont {Stinchcombe}(1981)}]{Stinchcombe1981}%
  \BibitemOpen
  \bibfield  {author} {\bibinfo {author} {\bibfnamefont {R.~B.}\ \bibnamefont
  {Stinchcombe}},\ }\bibfield  {title} {\bibinfo {title} {Diluted quantum
  transverse ising model},\ }\href
  {https://doi.org/10.1088/0022-3719/14/10/003} {\bibfield  {journal} {\bibinfo
   {journal} {Journal of Physics C: Solid State Physics}\ }\textbf {\bibinfo
  {volume} {14}},\ \bibinfo {pages} {L263–L267} (\bibinfo {year}
  {1981})}\BibitemShut {NoStop}%
\bibitem [{\citenamefont {Santos}(1982)}]{Santos1982}%
  \BibitemOpen
  \bibfield  {author} {\bibinfo {author} {\bibfnamefont {R.~R.~d.}\
  \bibnamefont {Santos}},\ }\bibfield  {title} {\bibinfo {title} {The pure and
  diluted quantum transverse ising model},\ }\href
  {https://doi.org/10.1088/0022-3719/15/14/020} {\bibfield  {journal} {\bibinfo
   {journal} {Journal of Physics C: Solid State Physics}\ }\textbf {\bibinfo
  {volume} {15}},\ \bibinfo {pages} {3141–3161} (\bibinfo {year}
  {1982})}\BibitemShut {NoStop}%
\bibitem [{\citenamefont {Fittipaldi}\ \emph {et~al.}(1985)\citenamefont
  {Fittipaldi}, \citenamefont {SáBarreto},\ and\ \citenamefont
  {Silva}}]{Fittipaldi1985}%
  \BibitemOpen
  \bibfield  {author} {\bibinfo {author} {\bibfnamefont {I.}~\bibnamefont
  {Fittipaldi}}, \bibinfo {author} {\bibfnamefont {F.}~\bibnamefont
  {SáBarreto}},\ and\ \bibinfo {author} {\bibfnamefont {P.}~\bibnamefont
  {Silva}},\ }\bibfield  {title} {\bibinfo {title} {Phase diagrams for the
  bond- and site-diluted transverse ising model},\ }\href
  {https://doi.org/10.1016/0378-4371(85)90134-7} {\bibfield  {journal}
  {\bibinfo  {journal} {Physica A: Statistical Mechanics and its Applications}\
  }\textbf {\bibinfo {volume} {131}},\ \bibinfo {pages} {599–611} (\bibinfo
  {year} {1985})}\BibitemShut {NoStop}%
\bibitem [{\citenamefont {Ikegami}\ \emph {et~al.}(1998)\citenamefont
  {Ikegami}, \citenamefont {Miyashita},\ and\ \citenamefont
  {Rieger}}]{Ikegami1998}%
  \BibitemOpen
  \bibfield  {author} {\bibinfo {author} {\bibfnamefont {T.}~\bibnamefont
  {Ikegami}}, \bibinfo {author} {\bibfnamefont {S.}~\bibnamefont {Miyashita}},\
  and\ \bibinfo {author} {\bibfnamefont {H.}~\bibnamefont {Rieger}},\
  }\bibfield  {title} {\bibinfo {title} {Griffiths-mccoy singularities in the
  transverse field ising model on the randomly diluted square lattice},\ }\href
  {https://doi.org/10.1143/jpsj.67.2671} {\bibfield  {journal} {\bibinfo
  {journal} {Journal of the Physical Society of Japan}\ }\textbf {\bibinfo
  {volume} {67}},\ \bibinfo {pages} {2671–2677} (\bibinfo {year}
  {1998})}\BibitemShut {NoStop}%
\bibitem [{\citenamefont {Thompson}\ and\ \citenamefont
  {Singh}(2019)}]{Thompson2019}%
  \BibitemOpen
  \bibfield  {author} {\bibinfo {author} {\bibfnamefont {F.}~\bibnamefont
  {Thompson}}\ and\ \bibinfo {author} {\bibfnamefont {R.~R.~P.}\ \bibnamefont
  {Singh}},\ }\bibfield  {title} {\bibinfo {title} {Griffiths-mccoy
  singularities in the dilute transverse-field ising model: A numerical linked
  cluster expansion study},\ }\bibfield  {journal} {\bibinfo  {journal}
  {Physical Review E}\ }\textbf {\bibinfo {volume} {99}},\ \href
  {https://doi.org/10.1103/physreve.99.032129} {10.1103/physreve.99.032129}
  (\bibinfo {year} {2019})\BibitemShut {NoStop}%
\bibitem [{\citenamefont {Kovács}\ and\ \citenamefont
  {Iglói}(2022)}]{Kovcs2022}%
  \BibitemOpen
  \bibfield  {author} {\bibinfo {author} {\bibfnamefont {I.~A.}\ \bibnamefont
  {Kovács}}\ and\ \bibinfo {author} {\bibfnamefont {F.}~\bibnamefont
  {Iglói}},\ }\bibfield  {title} {\bibinfo {title} {Geometry of rare regions
  behind griffiths singularities in random quantum magnets},\ }\bibfield
  {journal} {\bibinfo  {journal} {Scientific Reports}\ }\textbf {\bibinfo
  {volume} {12}},\ \href {https://doi.org/10.1038/s41598-022-05096-z}
  {10.1038/s41598-022-05096-z} (\bibinfo {year} {2022})\BibitemShut {NoStop}%
\bibitem [{\citenamefont {Senthil}\ and\ \citenamefont
  {Sachdev}(1996)}]{Senthil1996}%
  \BibitemOpen
  \bibfield  {author} {\bibinfo {author} {\bibfnamefont {T.}~\bibnamefont
  {Senthil}}\ and\ \bibinfo {author} {\bibfnamefont {S.}~\bibnamefont
  {Sachdev}},\ }\bibfield  {title} {\bibinfo {title} {Higher dimensional
  realizations of activated dynamic scaling at random quantum transitions},\
  }\href {https://doi.org/10.1103/physrevlett.77.5292} {\bibfield  {journal}
  {\bibinfo  {journal} {Physical Review Letters}\ }\textbf {\bibinfo {volume}
  {77}},\ \bibinfo {pages} {5292–5295} (\bibinfo {year} {1996})}\BibitemShut
  {NoStop}%
\bibitem [{\citenamefont {Sandvik}\ and\ \citenamefont
  {Kurkij\"arvi}(1991)}]{Sandvik1991}%
  \BibitemOpen
  \bibfield  {author} {\bibinfo {author} {\bibfnamefont {A.~W.}\ \bibnamefont
  {Sandvik}}\ and\ \bibinfo {author} {\bibfnamefont {J.}~\bibnamefont
  {Kurkij\"arvi}},\ }\bibfield  {title} {\bibinfo {title} {Quantum monte carlo
  simulation method for spin systems},\ }\href
  {https://doi.org/10.1103/PhysRevB.43.5950} {\bibfield  {journal} {\bibinfo
  {journal} {Phys. Rev. B}\ }\textbf {\bibinfo {volume} {43}},\ \bibinfo
  {pages} {5950} (\bibinfo {year} {1991})}\BibitemShut {NoStop}%
\bibitem [{\citenamefont {Sandvik}(1992)}]{Sandvik1992}%
  \BibitemOpen
  \bibfield  {author} {\bibinfo {author} {\bibfnamefont {A.~W.}\ \bibnamefont
  {Sandvik}},\ }\bibfield  {title} {\bibinfo {title} {A generalization of
  handscombs quantum monte carlo scheme-application to the 1d hubbard model},\
  }\href {https://doi.org/10.1088/0305-4470/25/13/017} {\bibfield  {journal}
  {\bibinfo  {journal} {Journal of Physics A: Mathematical and General}\
  }\textbf {\bibinfo {volume} {25}},\ \bibinfo {pages} {3667} (\bibinfo {year}
  {1992})}\BibitemShut {NoStop}%
\bibitem [{\citenamefont {Sandvik}(2003)}]{Sandvik2003}%
  \BibitemOpen
  \bibfield  {author} {\bibinfo {author} {\bibfnamefont {A.~W.}\ \bibnamefont
  {Sandvik}},\ }\bibfield  {title} {\bibinfo {title} {Stochastic series
  expansion method for quantum ising models with arbitrary interactions},\
  }\href {https://doi.org/10.1103/PhysRevE.68.056701} {\bibfield  {journal}
  {\bibinfo  {journal} {Phys. Rev. E}\ }\textbf {\bibinfo {volume} {68}},\
  \bibinfo {pages} {056701} (\bibinfo {year} {2003})}\BibitemShut {NoStop}%
\bibitem [{\citenamefont {Sandvik}\ \emph {et~al.}(2010)\citenamefont
  {Sandvik}, \citenamefont {Avella},\ and\ \citenamefont
  {Mancini}}]{Sandvik2010}%
  \BibitemOpen
  \bibfield  {author} {\bibinfo {author} {\bibfnamefont {A.~W.}\ \bibnamefont
  {Sandvik}}, \bibinfo {author} {\bibfnamefont {A.}~\bibnamefont {Avella}},\
  and\ \bibinfo {author} {\bibfnamefont {F.}~\bibnamefont {Mancini}},\
  }\bibfield  {title} {\bibinfo {title} {Computational studies of quantum spin
  systems},\ }in\ \href {https://doi.org/10.1063/1.3518900} {\emph {\bibinfo
  {booktitle} {{AIP} Conference Proceedings}}}\ (\bibinfo  {publisher}
  {{AIP}},\ \bibinfo {year} {2010})\BibitemShut {NoStop}%
\bibitem [{\citenamefont {Pelissetto}\ and\ \citenamefont
  {Vicari}(2002)}]{Pelissetto2002}%
  \BibitemOpen
  \bibfield  {author} {\bibinfo {author} {\bibfnamefont {A.}~\bibnamefont
  {Pelissetto}}\ and\ \bibinfo {author} {\bibfnamefont {E.}~\bibnamefont
  {Vicari}},\ }\bibfield  {title} {\bibinfo {title} {Critical phenomena and
  renormalization-group theory},\ }\href
  {https://doi.org/10.1016/s0370-1573(02)00219-3} {\bibfield  {journal}
  {\bibinfo  {journal} {Physics Reports}\ }\textbf {\bibinfo {volume} {368}},\
  \bibinfo {pages} {549–727} (\bibinfo {year} {2002})}\BibitemShut {NoStop}%
\bibitem [{\citenamefont {Sachdev}(2011)}]{Sachdev2007}%
  \BibitemOpen
  \bibfield  {author} {\bibinfo {author} {\bibfnamefont {S.}~\bibnamefont
  {Sachdev}},\ }\href {https://doi.org/10.1017/CBO9780511973765} {\emph
  {\bibinfo {title} {Quantum Phase Transitions}}},\ \bibinfo {edition} {2nd}\
  ed.\ (\bibinfo  {publisher} {Cambridge University Press},\ \bibinfo {year}
  {2011})\BibitemShut {NoStop}%
\bibitem [{\citenamefont {Kos}\ \emph {et~al.}(2016)\citenamefont {Kos},
  \citenamefont {Poland}, \citenamefont {Simmons-Duffin},\ and\ \citenamefont
  {Vichi}}]{Kos2016}%
  \BibitemOpen
  \bibfield  {author} {\bibinfo {author} {\bibfnamefont {F.}~\bibnamefont
  {Kos}}, \bibinfo {author} {\bibfnamefont {D.}~\bibnamefont {Poland}},
  \bibinfo {author} {\bibfnamefont {D.}~\bibnamefont {Simmons-Duffin}},\ and\
  \bibinfo {author} {\bibfnamefont {A.}~\bibnamefont {Vichi}},\ }\bibfield
  {title} {\bibinfo {title} {Precision islands in the ising and o(n ) models},\
  }\bibfield  {journal} {\bibinfo  {journal} {Journal of High Energy Physics}\
  }\textbf {\bibinfo {volume} {2016}},\ \href
  {https://doi.org/10.1007/jhep08(2016)036} {10.1007/jhep08(2016)036} (\bibinfo
  {year} {2016})\BibitemShut {NoStop}%
\bibitem [{\citenamefont {Bollobás}\ and\ \citenamefont
  {Riordan}(2005)}]{Bollobs2005}%
  \BibitemOpen
  \bibfield  {author} {\bibinfo {author} {\bibfnamefont {B.}~\bibnamefont
  {Bollobás}}\ and\ \bibinfo {author} {\bibfnamefont {O.}~\bibnamefont
  {Riordan}},\ }\bibfield  {title} {\bibinfo {title} {The critical probability
  for random voronoi percolation in the plane is 1/2},\ }\href
  {https://doi.org/10.1007/s00440-005-0490-z} {\bibfield  {journal} {\bibinfo
  {journal} {Probability Theory and Related Fields}\ }\textbf {\bibinfo
  {volume} {136}},\ \bibinfo {pages} {417–468} (\bibinfo {year}
  {2005})}\BibitemShut {NoStop}%
\bibitem [{\citenamefont {Newman}\ and\ \citenamefont
  {Ziff}(2000)}]{Newman2000}%
  \BibitemOpen
  \bibfield  {author} {\bibinfo {author} {\bibfnamefont {M.~E.~J.}\
  \bibnamefont {Newman}}\ and\ \bibinfo {author} {\bibfnamefont {R.~M.}\
  \bibnamefont {Ziff}},\ }\bibfield  {title} {\bibinfo {title} {Efficient monte
  carlo algorithm and high-precision results for percolation},\ }\href
  {https://doi.org/10.1103/physrevlett.85.4104} {\bibfield  {journal} {\bibinfo
   {journal} {Physical Review Letters}\ }\textbf {\bibinfo {volume} {85}},\
  \bibinfo {pages} {4104–4107} (\bibinfo {year} {2000})}\BibitemShut
  {NoStop}%
\bibitem [{\citenamefont {Nijs}(1979)}]{Nijs1979}%
  \BibitemOpen
  \bibfield  {author} {\bibinfo {author} {\bibfnamefont {M.~P. M.~d.}\
  \bibnamefont {Nijs}},\ }\bibfield  {title} {\bibinfo {title} {A relation
  between the temperature exponents of the eight-vertex and q-state potts
  model},\ }\href {https://doi.org/10.1088/0305-4470/12/10/030} {\bibfield
  {journal} {\bibinfo  {journal} {Journal of Physics A: Mathematical and
  General}\ }\textbf {\bibinfo {volume} {12}},\ \bibinfo {pages} {1857–1868}
  (\bibinfo {year} {1979})}\BibitemShut {NoStop}%
\bibitem [{\citenamefont {Pearson}(1980)}]{Pearson1980}%
  \BibitemOpen
  \bibfield  {author} {\bibinfo {author} {\bibfnamefont {R.~B.}\ \bibnamefont
  {Pearson}},\ }\bibfield  {title} {\bibinfo {title} {Conjecture for the
  extended potts model magnetic eigenvalue},\ }\href
  {https://doi.org/10.1103/physrevb.22.2579} {\bibfield  {journal} {\bibinfo
  {journal} {Physical Review B}\ }\textbf {\bibinfo {volume} {22}},\ \bibinfo
  {pages} {2579–2580} (\bibinfo {year} {1980})}\BibitemShut {NoStop}%
\bibitem [{\citenamefont {Sokolov}(1986)}]{Sokolov1986}%
  \BibitemOpen
  \bibfield  {author} {\bibinfo {author} {\bibfnamefont {I.~M.}\ \bibnamefont
  {Sokolov}},\ }\bibfield  {title} {\bibinfo {title} {Dimensionalities and
  other geometric critical exponents in percolation theory},\ }\href
  {https://doi.org/10.1070/pu1986v029n10abeh003526} {\bibfield  {journal}
  {\bibinfo  {journal} {Soviet Physics Uspekhi}\ }\textbf {\bibinfo {volume}
  {29}},\ \bibinfo {pages} {924–945} (\bibinfo {year} {1986})}\BibitemShut
  {NoStop}%
\bibitem [{\citenamefont {Stauffer}\ and\ \citenamefont
  {Aharony}(2018)}]{Stauffer2018}%
  \BibitemOpen
  \bibfield  {author} {\bibinfo {author} {\bibfnamefont {D.}~\bibnamefont
  {Stauffer}}\ and\ \bibinfo {author} {\bibfnamefont {A.}~\bibnamefont
  {Aharony}},\ }\href {https://doi.org/10.1201/9781315274386} {\emph {\bibinfo
  {title} {Introduction To Percolation Theory}}}\ (\bibinfo  {publisher}
  {Taylor \& Francis},\ \bibinfo {year} {2018})\BibitemShut {NoStop}%
\bibitem [{\citenamefont {Choi}\ and\ \citenamefont {Baek}(2023)}]{Choi2023}%
  \BibitemOpen
  \bibfield  {author} {\bibinfo {author} {\bibfnamefont {J.}~\bibnamefont
  {Choi}}\ and\ \bibinfo {author} {\bibfnamefont {S.~K.}\ \bibnamefont
  {Baek}},\ }\bibfield  {title} {\bibinfo {title} {Finite-size scaling analysis
  of the two-dimensional random transverse-field ising ferromagnet},\ }\href
  {https://doi.org/10.1103/PhysRevB.108.144204} {\bibfield  {journal} {\bibinfo
   {journal} {Phys. Rev. B}\ }\textbf {\bibinfo {volume} {108}},\ \bibinfo
  {pages} {144204} (\bibinfo {year} {2023})}\BibitemShut {NoStop}%
\bibitem [{\citenamefont {Kovács}(2022)}]{Kovacs2022}%
  \BibitemOpen
  \bibfield  {author} {\bibinfo {author} {\bibfnamefont {I.~A.}\ \bibnamefont
  {Kovács}},\ }\bibfield  {title} {\bibinfo {title} {Quantum multicritical
  point in the two- and three-dimensional random transverse-field ising
  model},\ }\bibfield  {journal} {\bibinfo  {journal} {Physical Review
  Research}\ }\textbf {\bibinfo {volume} {4}},\ \href
  {https://doi.org/10.1103/physrevresearch.4.013072}
  {10.1103/physrevresearch.4.013072} (\bibinfo {year} {2022})\BibitemShut
  {NoStop}%
\bibitem [{\citenamefont {Adelhardt}\ \emph {et~al.}(2024)\citenamefont
  {Adelhardt}, \citenamefont {Koziol}, \citenamefont {Langheld},\ and\
  \citenamefont {Schmidt}}]{Adelhardt2024}%
  \BibitemOpen
  \bibfield  {author} {\bibinfo {author} {\bibfnamefont {P.}~\bibnamefont
  {Adelhardt}}, \bibinfo {author} {\bibfnamefont {J.~A.}\ \bibnamefont
  {Koziol}}, \bibinfo {author} {\bibfnamefont {A.}~\bibnamefont {Langheld}},\
  and\ \bibinfo {author} {\bibfnamefont {K.~P.}\ \bibnamefont {Schmidt}},\
  }\bibfield  {title} {\bibinfo {title} {Monte carlo based techniques for
  quantum magnets with long-range interactions},\ }\href
  {https://doi.org/10.3390/e26050401} {\bibfield  {journal} {\bibinfo
  {journal} {Entropy}\ }\textbf {\bibinfo {volume} {26}},\ \bibinfo {pages}
  {401} (\bibinfo {year} {2024})}\BibitemShut {NoStop}%
\bibitem [{\citenamefont {Sandvik}(2002)}]{Sandvik2002}%
  \BibitemOpen
  \bibfield  {author} {\bibinfo {author} {\bibfnamefont {A.~W.}\ \bibnamefont
  {Sandvik}},\ }\bibfield  {title} {\bibinfo {title} {Classical percolation
  transition in the diluted two-dimensional $s=\frac{1}{2}$ heisenberg
  antiferromagnet},\ }\href {https://doi.org/10.1103/PhysRevB.66.024418}
  {\bibfield  {journal} {\bibinfo  {journal} {Phys. Rev. B}\ }\textbf {\bibinfo
  {volume} {66}},\ \bibinfo {pages} {024418} (\bibinfo {year}
  {2002})}\BibitemShut {NoStop}%
\bibitem [{\citenamefont {Wilson}(1971{\natexlab{a}})}]{Wilson1970}%
  \BibitemOpen
  \bibfield  {author} {\bibinfo {author} {\bibfnamefont {K.~G.}\ \bibnamefont
  {Wilson}},\ }\bibfield  {title} {\bibinfo {title} {Renormalization group and
  critical phenomena. i. renormalization group and the kadanoff scaling
  picture},\ }\href {https://doi.org/10.1103/PhysRevB.4.3174} {\bibfield
  {journal} {\bibinfo  {journal} {Phys. Rev. B}\ }\textbf {\bibinfo {volume}
  {4}},\ \bibinfo {pages} {3174} (\bibinfo {year}
  {1971}{\natexlab{a}})}\BibitemShut {NoStop}%
\bibitem [{\citenamefont {Wilson}(1971{\natexlab{b}})}]{Wilson1971}%
  \BibitemOpen
  \bibfield  {author} {\bibinfo {author} {\bibfnamefont {K.~G.}\ \bibnamefont
  {Wilson}},\ }\bibfield  {title} {\bibinfo {title} {Renormalization group and
  critical phenomena. ii. phase-space cell analysis of critical behavior},\
  }\href {https://doi.org/10.1103/PhysRevB.4.3184} {\bibfield  {journal}
  {\bibinfo  {journal} {Phys. Rev. B}\ }\textbf {\bibinfo {volume} {4}},\
  \bibinfo {pages} {3184} (\bibinfo {year} {1971}{\natexlab{b}})}\BibitemShut
  {NoStop}%
\bibitem [{\citenamefont {Hankey}\ and\ \citenamefont
  {Stanley}(1972)}]{Hankey1972}%
  \BibitemOpen
  \bibfield  {author} {\bibinfo {author} {\bibfnamefont {A.}~\bibnamefont
  {Hankey}}\ and\ \bibinfo {author} {\bibfnamefont {H.~E.}\ \bibnamefont
  {Stanley}},\ }\bibfield  {title} {\bibinfo {title} {Systematic application of
  generalized homogeneous functions to static scaling, dynamic scaling, and
  universality},\ }\href {https://doi.org/10.1103/PhysRevB.6.3515} {\bibfield
  {journal} {\bibinfo  {journal} {Phys. Rev. B}\ }\textbf {\bibinfo {volume}
  {6}},\ \bibinfo {pages} {3515} (\bibinfo {year} {1972})}\BibitemShut
  {NoStop}%
\bibitem [{\citenamefont {Brézin}(1982)}]{Brzin1982}%
  \BibitemOpen
  \bibfield  {author} {\bibinfo {author} {\bibfnamefont {E.}~\bibnamefont
  {Brézin}},\ }\bibfield  {title} {\bibinfo {title} {An investigation of
  finite size scaling},\ }\href
  {https://doi.org/10.1051/jphys:0198200430101500} {\bibfield  {journal}
  {\bibinfo  {journal} {Journal de Physique}\ }\textbf {\bibinfo {volume}
  {43}},\ \bibinfo {pages} {15–22} (\bibinfo {year} {1982})}\BibitemShut
  {NoStop}%
\bibitem [{\citenamefont {Brézin}\ and\ \citenamefont
  {Zinn-Justin}(1985)}]{Brzin1985}%
  \BibitemOpen
  \bibfield  {author} {\bibinfo {author} {\bibfnamefont {E.}~\bibnamefont
  {Brézin}}\ and\ \bibinfo {author} {\bibfnamefont {J.}~\bibnamefont
  {Zinn-Justin}},\ }\bibfield  {title} {\bibinfo {title} {Finite size effects
  in phase transitions},\ }\href {https://doi.org/10.1016/0550-3213(85)90379-7}
  {\bibfield  {journal} {\bibinfo  {journal} {Nuclear Physics B}\ }\textbf
  {\bibinfo {volume} {257}},\ \bibinfo {pages} {867–893} (\bibinfo {year}
  {1985})}\BibitemShut {NoStop}%
\bibitem [{\citenamefont {Binder}(1987)}]{Binder1987}%
  \BibitemOpen
  \bibfield  {author} {\bibinfo {author} {\bibfnamefont {K.}~\bibnamefont
  {Binder}},\ }\bibfield  {title} {\bibinfo {title} {{Finite size effects on
  phase transitions}},\ }\href {https://doi.org/10.1080/00150198708227908}
  {\bibfield  {journal} {\bibinfo  {journal} {Ferroelectrics}\ }\textbf
  {\bibinfo {volume} {73}},\ \bibinfo {pages} {43} (\bibinfo {year}
  {1987})}\BibitemShut {NoStop}%
\bibitem [{\citenamefont {Fisher}\ and\ \citenamefont
  {Barber}(1972)}]{Fisher1972}%
  \BibitemOpen
  \bibfield  {author} {\bibinfo {author} {\bibfnamefont {M.~E.}\ \bibnamefont
  {Fisher}}\ and\ \bibinfo {author} {\bibfnamefont {M.~N.}\ \bibnamefont
  {Barber}},\ }\bibfield  {title} {\bibinfo {title} {Scaling theory for
  finite-size effects in the critical region},\ }\href
  {https://doi.org/10.1103/PhysRevLett.28.1516} {\bibfield  {journal} {\bibinfo
   {journal} {Phys. Rev. Lett.}\ }\textbf {\bibinfo {volume} {28}},\ \bibinfo
  {pages} {1516} (\bibinfo {year} {1972})}\BibitemShut {NoStop}%
\bibitem [{\citenamefont {Kirkpatrick}\ and\ \citenamefont
  {Belitz}(2015)}]{Kirkpatrick2015}%
  \BibitemOpen
  \bibfield  {author} {\bibinfo {author} {\bibfnamefont {T.~R.}\ \bibnamefont
  {Kirkpatrick}}\ and\ \bibinfo {author} {\bibfnamefont {D.}~\bibnamefont
  {Belitz}},\ }\bibfield  {title} {\bibinfo {title} {Exponent relations at
  quantum phase transitions with applications to metallic quantum
  ferromagnets},\ }\href {https://doi.org/10.1103/PhysRevB.91.214407}
  {\bibfield  {journal} {\bibinfo  {journal} {Phys. Rev. B}\ }\textbf {\bibinfo
  {volume} {91}},\ \bibinfo {pages} {214407} (\bibinfo {year}
  {2015})}\BibitemShut {NoStop}%
\bibitem [{\citenamefont {Lin}\ \emph {et~al.}(2017)\citenamefont {Lin},
  \citenamefont {Kao}, \citenamefont {Chen},\ and\ \citenamefont
  {Lin}}]{Lin2017}%
  \BibitemOpen
  \bibfield  {author} {\bibinfo {author} {\bibfnamefont {Y.-P.}\ \bibnamefont
  {Lin}}, \bibinfo {author} {\bibfnamefont {Y.-J.}\ \bibnamefont {Kao}},
  \bibinfo {author} {\bibfnamefont {P.}~\bibnamefont {Chen}},\ and\ \bibinfo
  {author} {\bibfnamefont {Y.-C.}\ \bibnamefont {Lin}},\ }\bibfield  {title}
  {\bibinfo {title} {Griffiths singularities in the random quantum ising
  antiferromagnet: A tree tensor network renormalization group study},\ }\href
  {https://doi.org/10.1103/PhysRevB.96.064427} {\bibfield  {journal} {\bibinfo
  {journal} {Phys. Rev. B}\ }\textbf {\bibinfo {volume} {96}},\ \bibinfo
  {pages} {064427} (\bibinfo {year} {2017})}\BibitemShut {NoStop}%
\bibitem [{\citenamefont {Krämer}\ \emph {et~al.}(2025)\citenamefont
  {Krämer}, \citenamefont {Hörmann},\ and\ \citenamefont
  {Schmidt}}]{RawData}%
  \BibitemOpen
  \bibfield  {author} {\bibinfo {author} {\bibfnamefont {C.}~\bibnamefont
  {Krämer}}, \bibinfo {author} {\bibfnamefont {M.}~\bibnamefont {Hörmann}},\
  and\ \bibinfo {author} {\bibfnamefont {K.~P.}\ \bibnamefont {Schmidt}},\
  }\href {https://doi.org/10.5281/zenodo.15774391} {\bibinfo {title}
  {{Supplementary data to "Quantum Monte Carlo study of the bond- and
  site-diluted transverse-field Ising model"}}} (\bibinfo {year}
  {2025})\BibitemShut {NoStop}%
\bibitem [{\citenamefont {Krämer}\ \emph
  {et~al.}(2024{\natexlab{b}})\citenamefont {Krämer}, \citenamefont {Koziol},
  \citenamefont {Langheld}, \citenamefont {Hörmann},\ and\ \citenamefont
  {Schmidt}}]{zenodo_kraemer2024}%
  \BibitemOpen
  \bibfield  {author} {\bibinfo {author} {\bibfnamefont {C.}~\bibnamefont
  {Krämer}}, \bibinfo {author} {\bibfnamefont {J.~A.}\ \bibnamefont {Koziol}},
  \bibinfo {author} {\bibfnamefont {A.}~\bibnamefont {Langheld}}, \bibinfo
  {author} {\bibfnamefont {M.}~\bibnamefont {Hörmann}},\ and\ \bibinfo
  {author} {\bibfnamefont {K.~P.}\ \bibnamefont {Schmidt}},\ }\href
  {https://doi.org/10.5281/zenodo.11401142} {\bibinfo {title} {{Supplementary
  data to "Quantum-critical properties of the one- and two-dimensional random
  transverse- field Ising model from large-scale quantum Monte Carlo
  simulations"}}} (\bibinfo {year} {2024}{\natexlab{b}})\BibitemShut {NoStop}%
\bibitem [{\citenamefont {Rieger}\ and\ \citenamefont
  {Young}(1996)}]{Rieger1996}%
  \BibitemOpen
  \bibfield  {author} {\bibinfo {author} {\bibfnamefont {H.}~\bibnamefont
  {Rieger}}\ and\ \bibinfo {author} {\bibfnamefont {A.~P.}\ \bibnamefont
  {Young}},\ }\bibfield  {title} {\bibinfo {title} {Griffiths singularities in
  the disordered phase of a quantum ising spin glass},\ }\href
  {https://doi.org/10.1103/PhysRevB.54.3328} {\bibfield  {journal} {\bibinfo
  {journal} {Phys. Rev. B}\ }\textbf {\bibinfo {volume} {54}},\ \bibinfo
  {pages} {3328} (\bibinfo {year} {1996})}\BibitemShut {NoStop}%
\bibitem [{\citenamefont {Pich}\ \emph {et~al.}(1998)\citenamefont {Pich},
  \citenamefont {Young}, \citenamefont {Rieger},\ and\ \citenamefont
  {Kawashima}}]{Pich1998}%
  \BibitemOpen
  \bibfield  {author} {\bibinfo {author} {\bibfnamefont {C.}~\bibnamefont
  {Pich}}, \bibinfo {author} {\bibfnamefont {A.~P.}\ \bibnamefont {Young}},
  \bibinfo {author} {\bibfnamefont {H.}~\bibnamefont {Rieger}},\ and\ \bibinfo
  {author} {\bibfnamefont {N.}~\bibnamefont {Kawashima}},\ }\bibfield  {title}
  {\bibinfo {title} {Critical behavior and griffiths-mccoy singularities in the
  two-dimensional random quantum ising ferromagnet},\ }\href
  {https://doi.org/10.1103/PhysRevLett.81.5916} {\bibfield  {journal} {\bibinfo
   {journal} {Phys. Rev. Lett.}\ }\textbf {\bibinfo {volume} {81}},\ \bibinfo
  {pages} {5916} (\bibinfo {year} {1998})}\BibitemShut {NoStop}%
\end{thebibliography}
